# SUPERNOVAE FROM ROTATING STARS

Georges MEYNET, André MAEDER

Geneva Observatory of Geneva University, 51 chemin des Maillettes, CH-1290 Versoix Geneva, Switzerland

Georges.Meynet@unige.ch, andre.maeder@unige.ch

**Abstract.** The present paper discusses the main physical effects produced by stellar rotation on presupernovae, as well as observations which confirm these effects and their consequences for presupernova models. Rotation critically influences the mass of the exploding cores, the mass and chemical composition of the envelopes and the types of supernovae, as well as the properties of the remnants and the chemical yields. In the formation of gamma-ray bursts, rotation and the properties of rotating stars appear as the key factor. In binaries, the interaction between axial rotation and tidal effects often leads to interesting and unexpected results. Rotation plays a key role in shaping the evolution and nucleosynthesis in massive stars with very low metallicities (metallicity below about the Small Magellanic Cloud metallicity down to Population III stars). At solar and higher metallicities, the effects of rotation compete with those of stellar winds. In close binaries, the synchronisation process can lock the star at a high rotation rate despite strong mass loss and thus both effects, rotation and stellar winds, have a strong impact. In conclusion, rotation is a key physical ingredient of the stellar models and of presupernova stages, and the evolution both of single stars and close binaries. Moreover, important effects are expected along the whole cosmic history.

## 1. INTRODUCTION

"Eppur si muove!" And yet it moves! This famous sentence, pronounced by Galileo Galilei about the motion of the Earth around the Sun, can also be applied to the rotational motion of stars around their axis. Often considered as a side effect in star evolution, rotation appears more and more to play a major role. All the outputs of stellar evolution are influenced by rotation: lifetimes, the size of convective cores, chemical surface enrichments, mass loss, the properties of presupernovae, chemical yields, the final stages,… All properties of binary evolution also depend on the axial rotation of the stars. For two or three decades, models and observations have demonstrated the need to account for stellar rotation. High resolution spectroscopy of fast-rotating massive stars has shown large enrichments of nitrogen, as well as for a few other elements. Observations in asteroseismology also demonstrate the occurrence of internal shears in stars, which likely are responsible for internal mixing of the elements.

## 2. THE MAIN PHYSICAL EFFECTS OF ROTATION

Let us examine the main physical effects of rotation, which may significantly influence the properties of presupernovae, as well as the nature of the explosion and of the remnants. The mixing effects produced by rotation have a relatively long timescale and thus they mainly act in the first evolutionary phases. Such effects must be carefully considered. The way they are treated often leads to large differences between authors.

## 2.1 The mechanical structure

Rotation would ideally demand a two or three-dimensional treatment. One-dimensional (1-D) models are, for now, the only ones able to follow the evolution of the structure of a star from formation to the supernova stage. In such models, the simplifying assumption of "shellular rotation" is often made. It assumes a rotation law $\Omega \sim$ constant on isobaric shells and depends to the first order on the distance r to the stellar centre (Zahn, 1992). A shellular rotation law results from the strong horizontal turbulence $D_h$ in differentially rotating stars, which imposes constancy of $\Omega$ on isobars, while in the vertical direction, the turbulence is weaker due to the stable density stratification.

The equations of stellar structure for such a rotation law in 1-D models have been obtained by Meynet & Maeder (1997) and have been extensively applied in stellar models up to the advanced stages. A rotating star also obeys the von Zeipel (1924) theorem that states that the flux at a given colatitude is proportional to the local effective gravity there. This theorem, also valid for shellular rotation, implies that the flux and thus the effective temperature decreases from the pole to the equator. The surface of a rotating star is thus inhomogeneous in effective gravity and temperature, which influences the emergent spectrum and mass loss.

## 2.2 The transport of chemical elements and angular momentum

A careful treatment of mixing is very necessary. Comparison of rotating models of presupernova (Sect. 3.1) shows that the main source of difference between authors lies in the way mixing is treated and in particular the inhibiting effects of the μ-gradients. Rotation drives internal currents and various turbulent motions, which may contribute to internal transport. Quite generally the transport of chemical elements, with an abundance $X_i$ in mass fraction, can be treated as a diffusion process obeying the Lagrangian equation at each level $M_r$ in the star:

$$\rho \left.\frac{\partial X_i}{\partial t}\right]_{M_r} = \frac{1}{r^2}\frac{\partial}{\partial r}\left(\rho r^2 D \frac{\partial X_i}{\partial r}\right).$$

.

There, D is the appropriate diffusion coefficient for the various instabilities considered. In a rotating star, the evolution of the angular velocity $\Omega(r,t)$ has to be followed at each level r. There is a loop of interactions: $\Omega(r,t)$ influences the transport of the angular momentum, which in turn also influences the angular velocity. The consistency of the whole loop has to be treated carefully (Maeder, 2009). The evolution of $\Omega(r,t)$ depends on both the vertical component $U_2$ of the meridional circulation velocity and on a diffusion coefficient D:

$$\rho \frac{\partial}{\partial t}(r^2 \overline{\Omega})_{M_r} = \frac{1}{5r^2}\frac{\partial}{\partial r}(\varrho r^4 \overline{\Omega} U_2(r)) + \frac{1}{r^2}\frac{\partial}{\partial r}\left(\varrho D r^4 \frac{\partial \overline{\Omega}}{\partial r}\right).$$

The first term on the right is an advective term resulting from the currents of meridional circulation. Ignoring this term, as is often done, is not a consistent choice. For the transport of angular momentum, the local conservation of the angular momentum limits the effects of the horizontal turbulence and $D_h$ does not play a significant role there. We see that the equation for chemical mixing only contains a term depending on D. The reason has been given by Zahn (1992). It rests on the fact that the horizontal turbulence interacts with circulation, so that for the chemical elements the circulation is demonstrated to be equivalent to a diffusion with a coefficient $D_{eff} = (r\ U_2)^2/(30\ D_h)$. Let us just note that the time scale of the meridional circulation behaves like the Kelvin-Helmholtz timescale divided by the ratio of the centrifugal force to gravity, so that meridional circulation significantly acts in long nuclear burning phases.

The main instability contributing to both transports is the shear instability, due to the velocity differences of adjacent layers. The instability criterion is the Richardson criterion, which says that an instability develops if the excess of energy in differential motions (excess with respect to the energy for an average velocity) is bigger than the work necessary to exchange matter vertically. It also depends on the gradient of composition, as well as on the horizontal turbulence. This is a particularly critical point. In this respect, we may note that the account of the horizontal turbulence weakens the inhibiting effects of the μ-gradients. Several other instabilities may also play a role for the transport of chemical elements and angular momentum, in particular the thermohaline mixing, the Rayleigh-Taylor, the Solberg-Hoiland and GSF instabilities (Maeder, 2009). We note in particular, that the Rayleigh-Taylor instability could intervene in the pre-supernova stages when the Ω–gradient is very steep at the edge of the stellar core.

### 2.3 The magnetic field

Degenerate objects in the final stages of stellar evolution have strong magnetic fields. Conservation of the magnetic flux of Main Sequence stars with fields in the range of 10 G to 1 kG would lead to neutron stars with magnetic fields in the range of $10^{11}$ to $10^{13}$ G and to white dwarfs with fields in the range of $10^5$ to $10^7$ G. This is in agreement with the fields of classical pulsars, which is of the order of $10^{12}$ G, and with those of magnetic white dwarfs. These orders of magnitude are consistent with the idea of fossil fields. However, there is room also for the existence of an internal dynamo in rotating stars. The generation of magnetic fields in convective regions has been mainly studied in the case of the solar dynamo. The existence of a dynamo in rotating radiative zones is still a major uncertainty. The main effect of a magnetic field in stars is to create an efficient coupling between adjacent regions. A viscosity coefficient can be associated to this coupling (Maeder and Meynet, 2012). It leads to a rotation law not far from solid body rotation, with deviations that still allow some chemical transport.

### 2.4 Rotation in binary stars

Tidal interactions play an important role in close binary evolution (de Mink et al., 2009), the extreme case of tidal interaction being the mass transfer between the two components. Depending on the values of the orbital period and of the rotation velocities, the tides may spin up or spin down the axial rotation of the components of the system. In both cases, tidal interaction produces a high internal differential rotation. The resulting shear mixing is very efficient, producing high N- and He-enrichments at the stellar surface (Song et al., 2013). Evolutionary tracks and lifetimes are modified. The tidal and rotational interactions may be so high that due to mixing the primary star becomes chemically homogenous. Such an evolution typically occurs when the rotation velocities are higher than 250 km s$^{-1}$ (this velocity limit is for models with a strong coupling mediated by an internal magnetic field). If this happens, the stellar radius no longer increases during evolution and, if the stars are compact enough (especially at lower metallicity Z), they may avoid Roche lobe overflow (RLOF) and mass transfer. The avoidance of RLOF is favoured in high mass stars (initial M> 35 $M_{Sun}$) with lower Z and shorter orbital periods (Song et al., 2015). This scenario of homogeneous evolution in a binary system is of importance in the context of the formation of binary black holes.

### 2.5    Observations of rotational effects

The stellar models predict that surface enrichment in products of CNO burning, observable for example by the N/H, or N/C ratios, increases with stellar mass M and rotation velocity v, because mixing gets stronger. The enrichments also increase with the age t, since more and more new nuclear products reach the stellar surface. The comparison of Figs. 1 and 2 illustrates the predicted effects of rotation for different masses and evolutionary stages. The enrichments are also stronger at lower Z, shear mixing being favoured in more compact stars. Tidal interactions, as seen above, may influence the mixing. Thus, chemical enrichments are a multivariate function (Maeder, 2009), e.g.

$$(N/H) = f(M, v, t, Z, binarity, …).$$

Neglect of the richness of the rotational effects by some authors who just concentrate on a simple relation of the form (N/H) = f(v) has led to a decade of useless debate.

The He-enrichments at the surface occur much more slowly than the N-enrichments, since the internal He-gradient is much weaker that the N-gradient. Careful analyses have well confirmed the predicted effects. A tight trend in the observed N/C vs. N/O ratios and the build-up of helium is found from main-sequence to supergiant stars (Przybilla et al., 2010). A careful spectroscopic survey of galactic O-stars by Martins et al. (2015) confirms many predicted effects of rotation.  Fig. 3 well illustrates the age effect on the chemical enrichments. A clear trend of stronger N-enrichments in more evolved stars is observed. The trend of greater mixing in more massive objects is present among both dwarfs and supergiants. 90% of the sample stars are well accounted by the rotational effects predicted by the models of Ekström et al. (2012). The evidence of greater mixing at lower Z is supported by the evidence of a production of primary nitrogen in low Z stars (Meynet & Maeder, 2002). Observations of the ratio $^{12}C/^{13}C$ in low Z stars, as well as other observations (Sect. 5), also support these conclusions (Chiappini et al., 2008). As to the magnetic effects in rotating stars, a variety of N-enrichments is observed in magnetic OB stars so that it is unclear for now if mixing is systematically driven by magnetic fields (see Fig. 3).

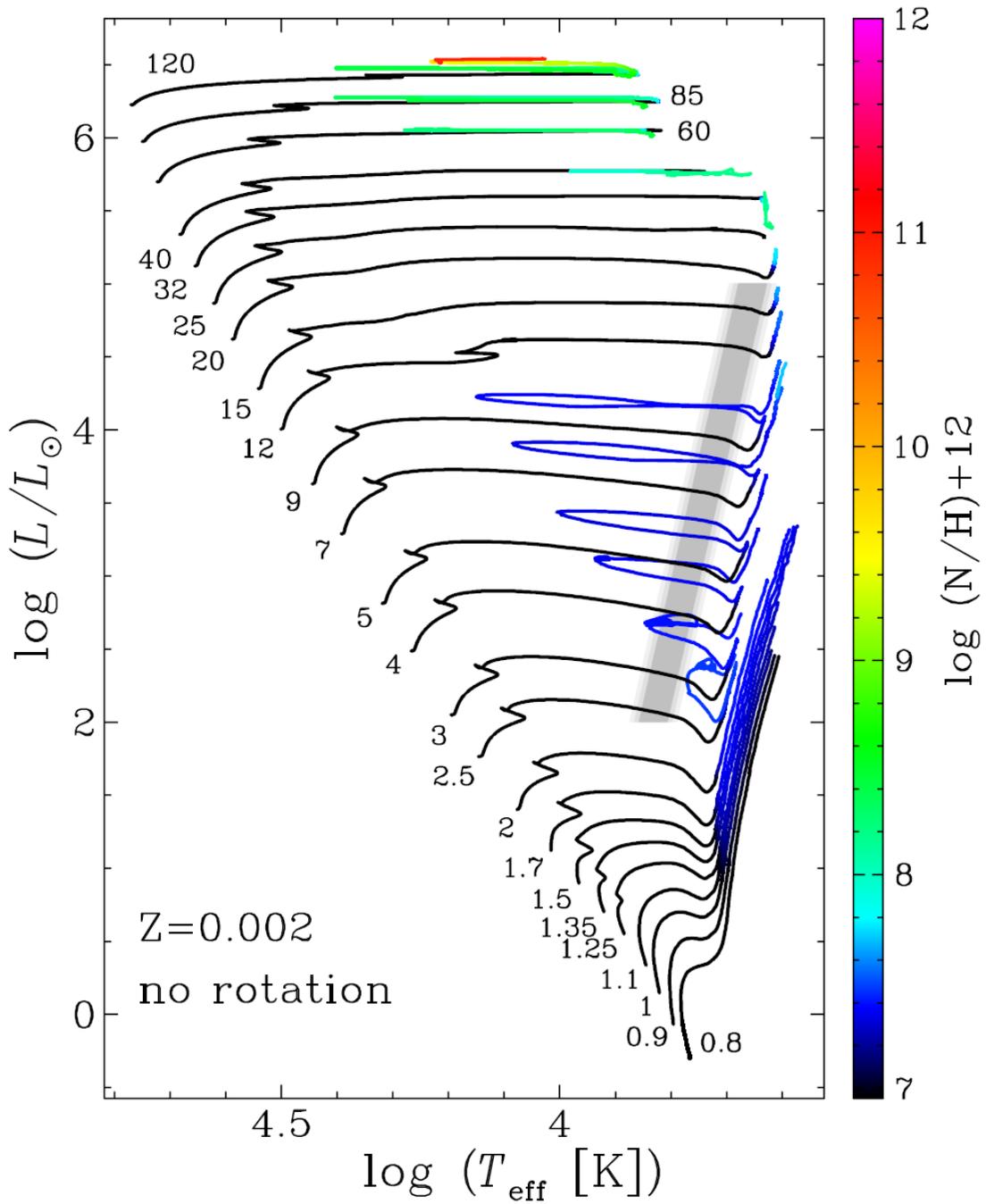

**FIG. 1**

HR diagram for non-rotating models at Z=0.002. The colour scale indicates the surface number abundance of nitrogen on a log scale where the H-abundance is 12. (From Georgy et al. 2013).

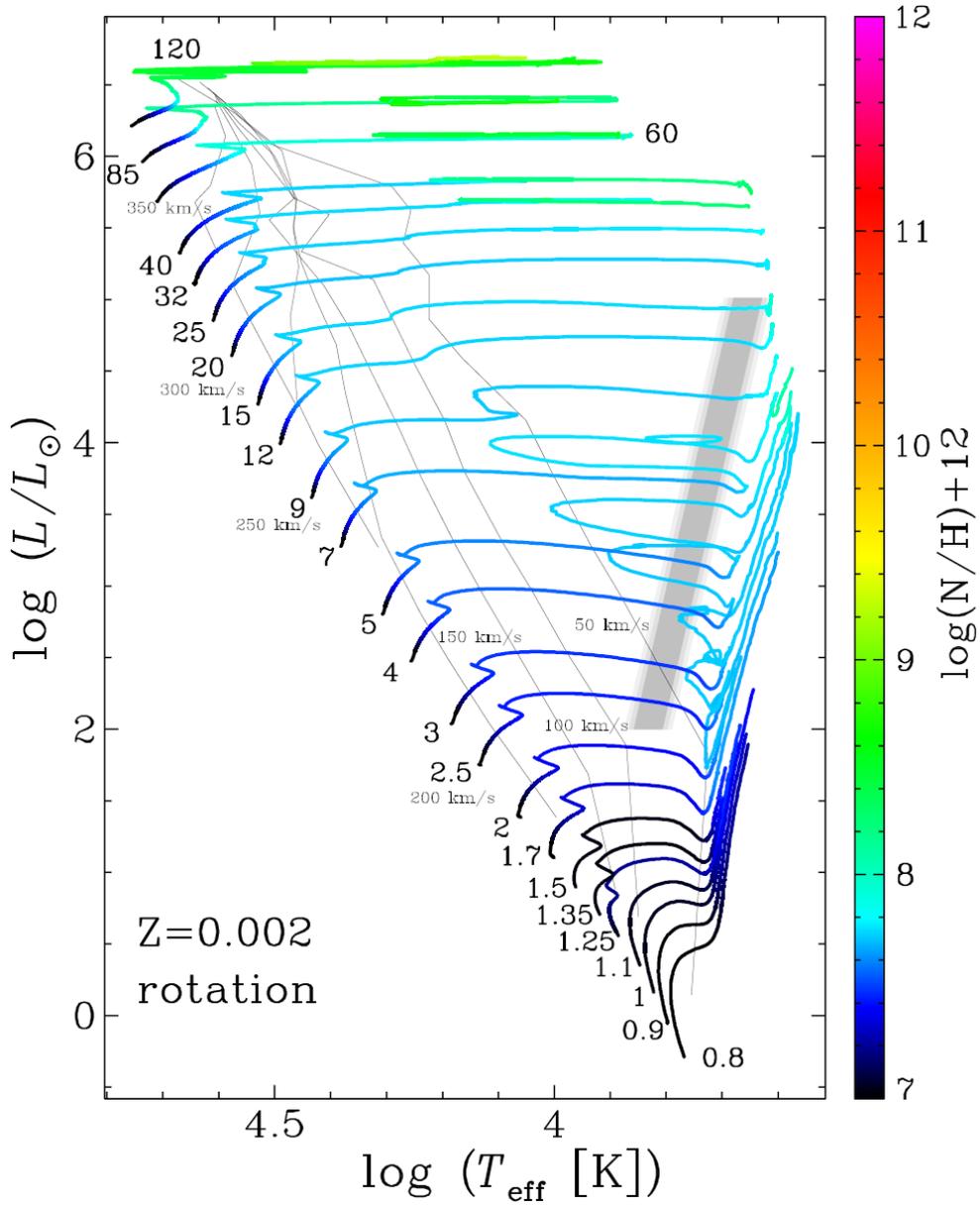

**FIG. 2**

HR diagram for rotating models at Z=0.002. Lines of iso-velocities are indicated. Same remarks as for Fig. 1. (From Georgy et al. 2013).

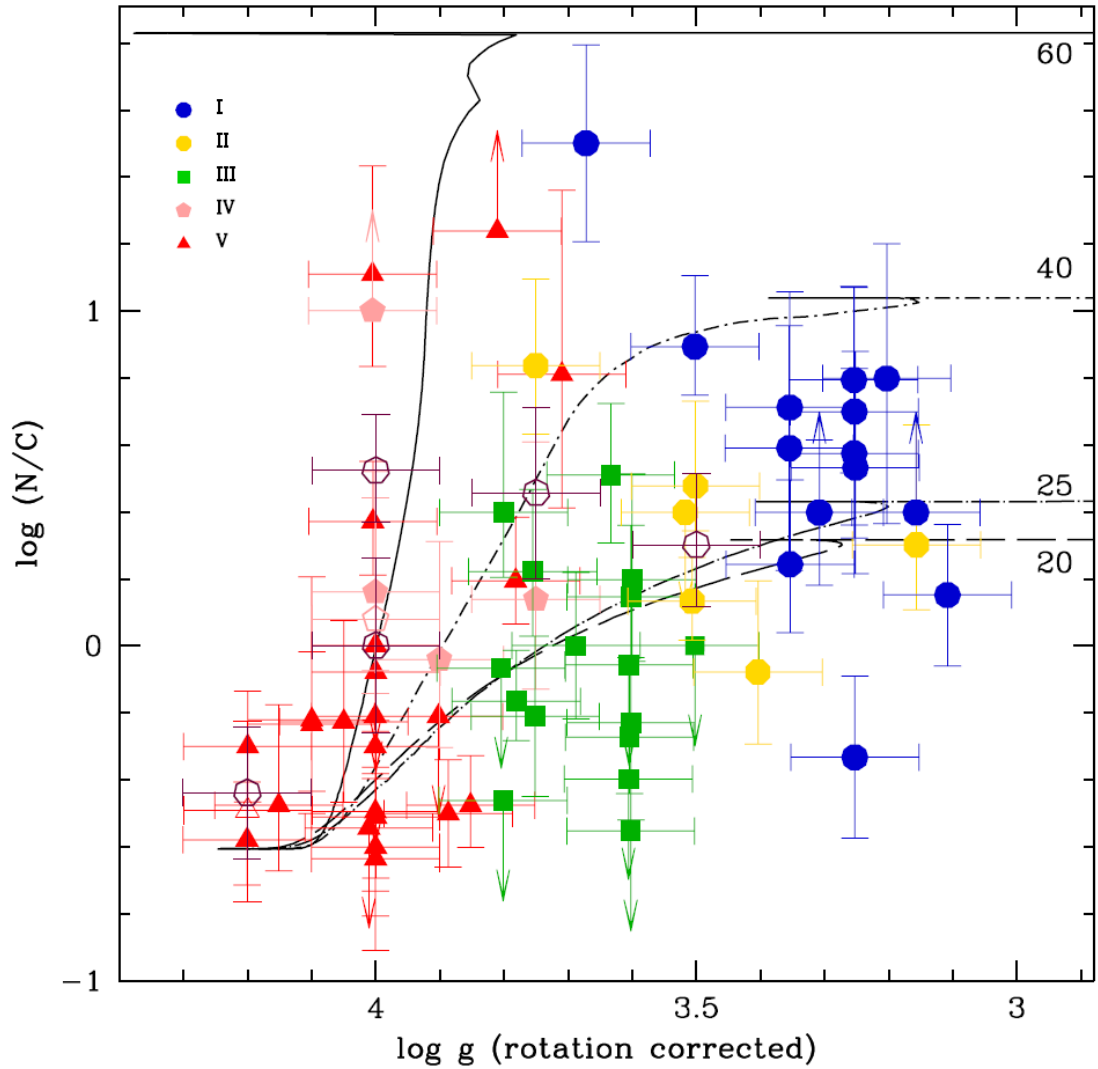

**FIG. 3**

Log (N/C) vs. log (gravity) for a sample of galactic OB stars. Red triangles are dwarfs, pink pentagons are subgiants, green square are giants, yellow octogons are bright giants, blue circles are supergiants. Open symbols represent magnetic stars (with magenta heptagons being Of?p stars). Evolutionary tracks with rotation are from Ekström et al. (2012). Typical error bars are given. (From Martins et al. 2015).

## 3. EVOLUTIONARY EFFECTS OF ROTATION ON THE PRE-SUPERNOVAE INTERNAL STRUCTURE

3.1 Stellar cores: masses and composition.

Most of the differences between the pre-supernova structures of rotating and non-rotating stellar models are shaped during the first phases of evolution. Lifetimes in the advanced stages are too short to allow most rotational instabilities to influence significantly the late stages. Fig. 4 (Hirschi et al., 2004) shows various critical masses as a function of the initial masses for non-rotating and rotating massive stars with Z=0.02 calculated up to the end of Si-core burning. The final masses for stars with rotation are smaller than for non-rotating stars, since rotation favours higher mass loss rates. The final masses of the most massive stars show a convergence towards about the same limit close to 15 Mo, because the mass loss rates in the WR stages depend on the actual mass.

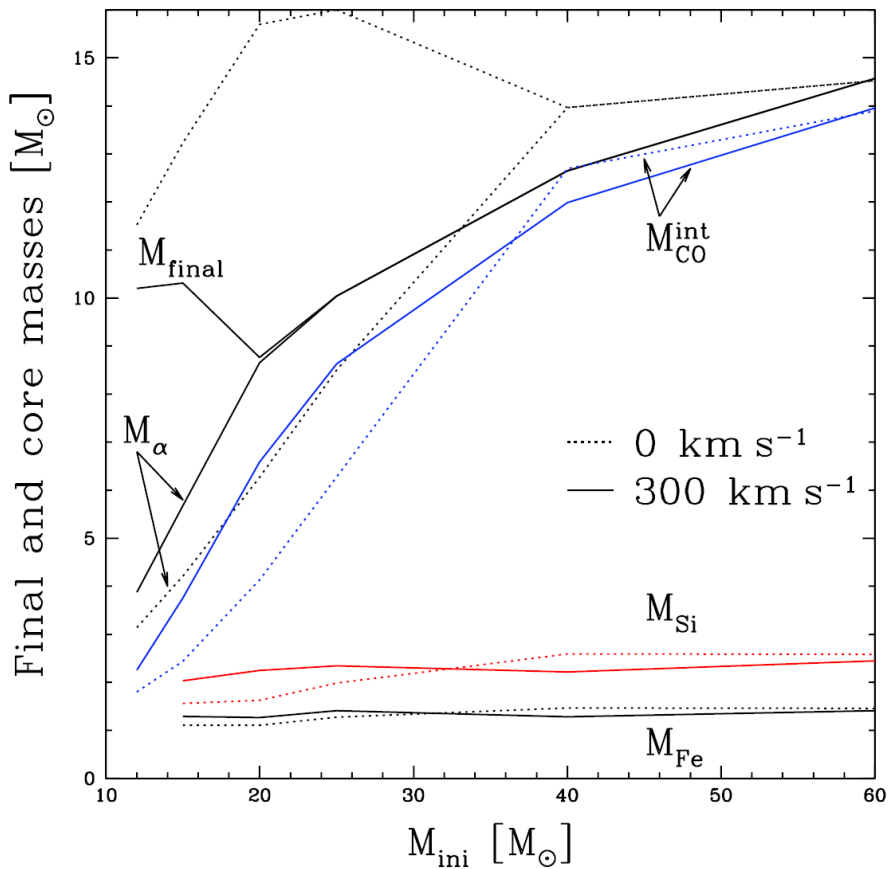

# FIG. 4

Different core masses at the end of core Si-burning as a function of the initial mass for two initial rotation velocities. The initial metallicity is Z=0.02. (From Hirschi et al. 2004).

$M_\alpha$ and $M_{CO}$ respectively are the mass of the He- and CO-cores, these masses essentially determine the types of supernova events. For initial masses M<~30 $M_{Sun}$, rotation significantly increases these two core masses due to internal mixing. For $M$ >~30 $M_{Sun}$, a rotating star enters at an earlier stage into the Wolf-Rayet (WR) phase and spends a longer time in this phase, which is characterised by heavy mass loss. (For a more extended definition of WR stars see Section 4.4). This results in smaller values of $M_\alpha$ and $M_{CO}$ for rotating stars at the pre-supernova stage. We also see that $M_\alpha$ in the rotating case is equal to the final masses for initial masses above 20 $M_{Sun}$, showing that these objects are WR stars. In the absence of rotation, the lower mass limit for WR stars is 25 $M_{Sun}$ at Z=0.014 (Georgy et al. 2012). The masses of the Si and Fe cores show the same behaviour as for $M_\alpha$ and $M_{CO}$, but the effects of rotation are smaller. As to the composition of the different onion-skin layers at the presupernova stage, the differences produced by rotation are very small in the deep interior, while they are larger in the outer layers, especially for the H-rich envelope (if there is one), and the He- and C-rich layers.

Comparisons between the results of various authors concerning the core masses have been made by Hirschi et al. (2005a) for stars with initial masses between 15 and 25 $M_{Sun}$. For $M_\alpha$ the differences are small (<10%). For $M_{CO}$, even for models without rotation, they almost reach a factor of two at 25 $M_{Sun}$, as a consequence of two effects:

  i. The different convective criteria, Schwarzschild, Ledoux, the way overshooting and semiconvection are treated.
  ii. The different reaction rates used for the critical $^{12}C(\alpha,\gamma)^{16}O$ reaction considerably influence the He- and C-burning phases.

In the case of rotating models, it appears that the critical source of differences between authors lies in the way the inhibiting effects of the μ-gradients on the mixing processes are accounted for. This confirms that the treatment of the mixing processes during the MS and He-burning phase is a critical point for the pre-supernova models.

The effects of rotation on the chemical yields, including both the supernova contribution and the winds of massive stars have been calculated by Hirschi et al. (2005a) for models in the range of 12 to 60 $M_{Sun}$. Below ~30 $M_{Sun}$, rotation increases the total yields in metals, Z, and in particular the yields of carbon and oxygen by a factor between 1.5 and 2.5. As a rule of thumb, the yields of a rotating 20 $M_{Sun}$ star are similar to the yields of a non-rotating 30 $M_{Sun}$ star, at least for the light α-elements considered. For very massive stars (~60 $M_{Sun}$), rotation increases the yield of helium but does not significantly affect the yields of heavy elements.

3.2     Stellar cores: angular momentum and relation with the rotation of pulsars.

The evolution of the angular velocity Ω is governed by the various mixing processes, in particular by meridional circulation, as well as by convection, contraction and expansion. As to the mass loss by stellar winds, it removes angular momentum from the stellar surface. Fig. 5 (Hirschi et al. 2004) shows the evolution of Ω inside the 25 $M_{Sun}$ rotating model from the beginning of the Main Sequence (MS), where Ω is considered as a constant, up to the end of the core Si-burning phase. During the MS phase, Ω decreases in the outer layers, which expand and lose mass. In the red supergiant phase, Ω increases in the core due to contraction, while the expansion of the outer layers and the mass loss lower the external velocity very much. The entire envelope is lost during the He-burning phase. After this phase, the nuclear timescales are so short that the mild mixing processes cease to be efficient. The evolution of $\Omega$ is essentially determined by convection, which flattens the $\Omega$-profile and by the local conservation of the angular momentum. The strong central contraction continues to increase $\Omega$ during the successive phases of nuclear burning, so that $\Omega$ reaches values of the order of 1 s$^{-1}$ at the end of Si-burning, a value still much below the break-up limit.

While the angular velocity $\Omega$ may change during evolution by 4 orders of magnitude or more (see Fig. 5), the evolution of the specific angular momentum $j(r) = (2/3) \Omega(r) r^2$ is completely different, as shown by Fig. 6, which refers to the same 25 M$_{Sun}$ model as in Fig. 5. The specific angular momentum at a given location in the star varies only by the effect of transport, while contraction or expansion let it unmodified. We see that during the MS phase $j(r)$ is reduced everywhere in the star, with a higher reduction in the convective core. There is some decrease until the end of the He-burning phase. Then, there is no global reduction, but the formation of teeth by convective zones, which, by making $\Omega$ constant, transport angular momentum from the inner to the outer part of a convective zone. The result is that the global picture of the angular momentum at the end of Si-burning is almost the same as at the end of He-burning (Heger et al. 2000, Hirschi et al. 2004). For the 25 M$_{Sun}$ model, the angular momentum of the 2.7 M$_{Sun}$ remnant is $2.15 \times 10^{50}$ g cm$^2$ s$^{-1}$ at the end of He-burning, while it is $1.63 \times 10^{50}$ g cm$^2$ s$^{-1}$ at the end of Si-burning, a loss of only 24% while $\Omega$ has changed by 4 orders of magnitude. This change is to be compared to the reduction of $j(r)$ by a factor about 5 on average between the ZAMS and the end of He-burning. This confirms our repeated statement about the importance of a careful treatment of all transport mechanisms, both for the chemical elements and the angular momentum.

It is interesting to compare the rotation of the stellar cores in the pre-supernova stages with the rotation of pulsars (Heger et al. 2005). We found above a typical angular velocity $\Omega \sim 1$ s$^{-1}$ in the core for a density $\rho$ of about $10^8$ g cm$^{-3}$ in the presupernova stage. For a mass element contracting with angular momentum conservation, the angular velocity varies like $\Omega \sim r^{-2} \sim \rho^{2/3}$. Neutron stars have a typical density above $10^{14}$ g cm$^{-3}$; thus they should spin with initial period of about 0.1 ms. The periods at birth are likely above 10 ms (we do not consider the so-called millisecond pulsars, usually interpreted as having been accelerated by mass accretion from a companion). This is about two orders of magnitude slower than predicted.

One may wonder at what evolutionary stage occurs the braking implied by these values. Among various possibilities considered, the effect of a magnetic field appears to be a powerful one. It may act in at least two ways. First, by imposing a solid body rotation, a lot of angular momentum is evacuated from stellar cores, so that the corresponding pulsar rotation may already be within an order of a magnitude of the observed one. Second, surface magnetic fields in rotating early type stars with stellar winds produce an efficient coupling between the star and the winds, with an important removal of angular momentum (Meynet et al., 2011). Even an initial field of 100 G produces a reduction of rotation by one order of magnitude, in addition to the effect of solid body rotation, while for 1 kG the additional reduction of the pulsar rotation would reach four orders of magnitude.

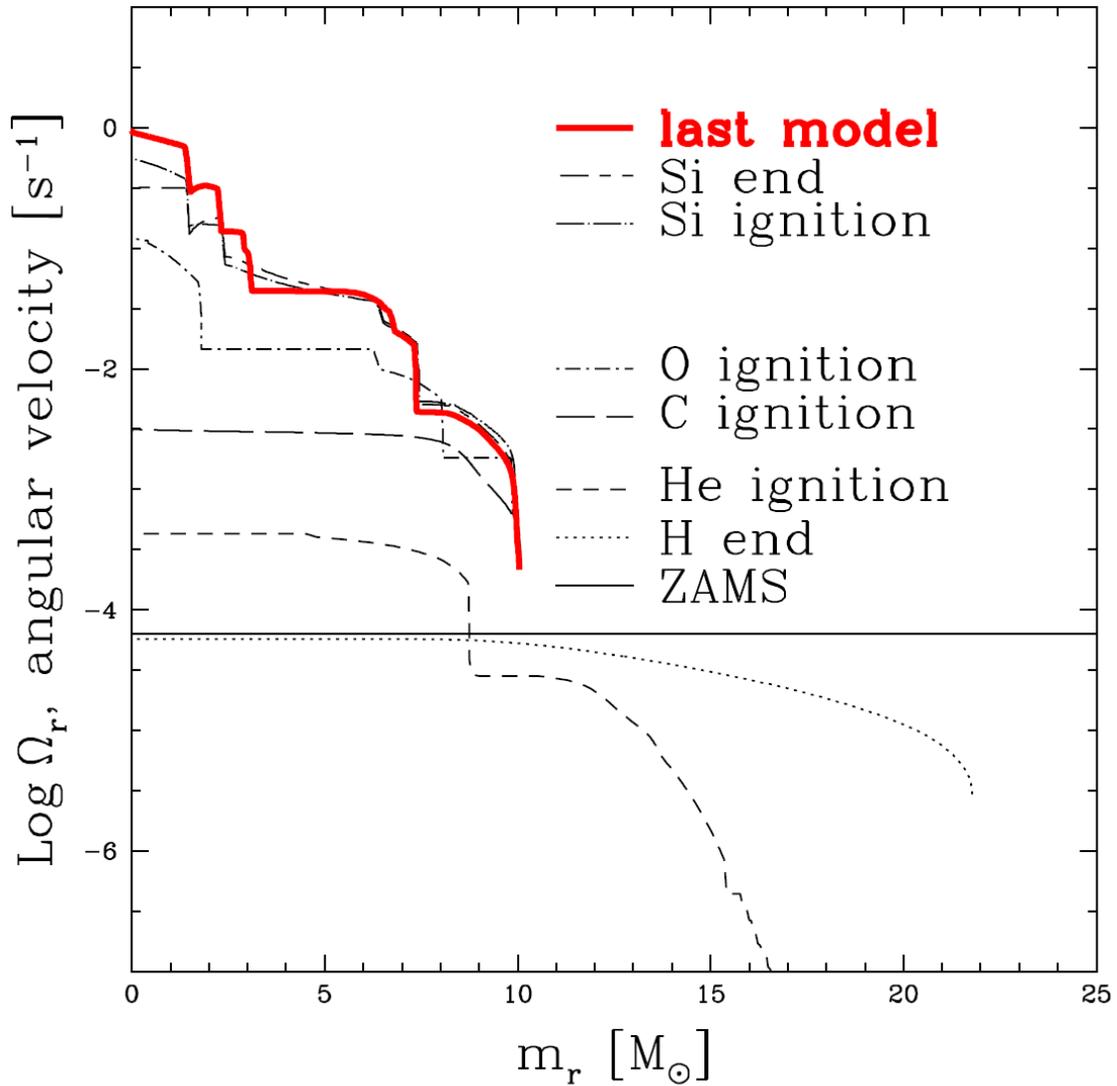

# FIG. 5

The internal distribution of the angular velocity at various evolutionary stages as a function of the Lagrangian mass coordinates inside a 25 Mo model with an initial rotation velocity of 300 km s$^{-1}$ and metallicity Z=0.02. An overshooting distance equal to 10% of the local pressure scale height has been applied. (From Hirschi et al. 2004).

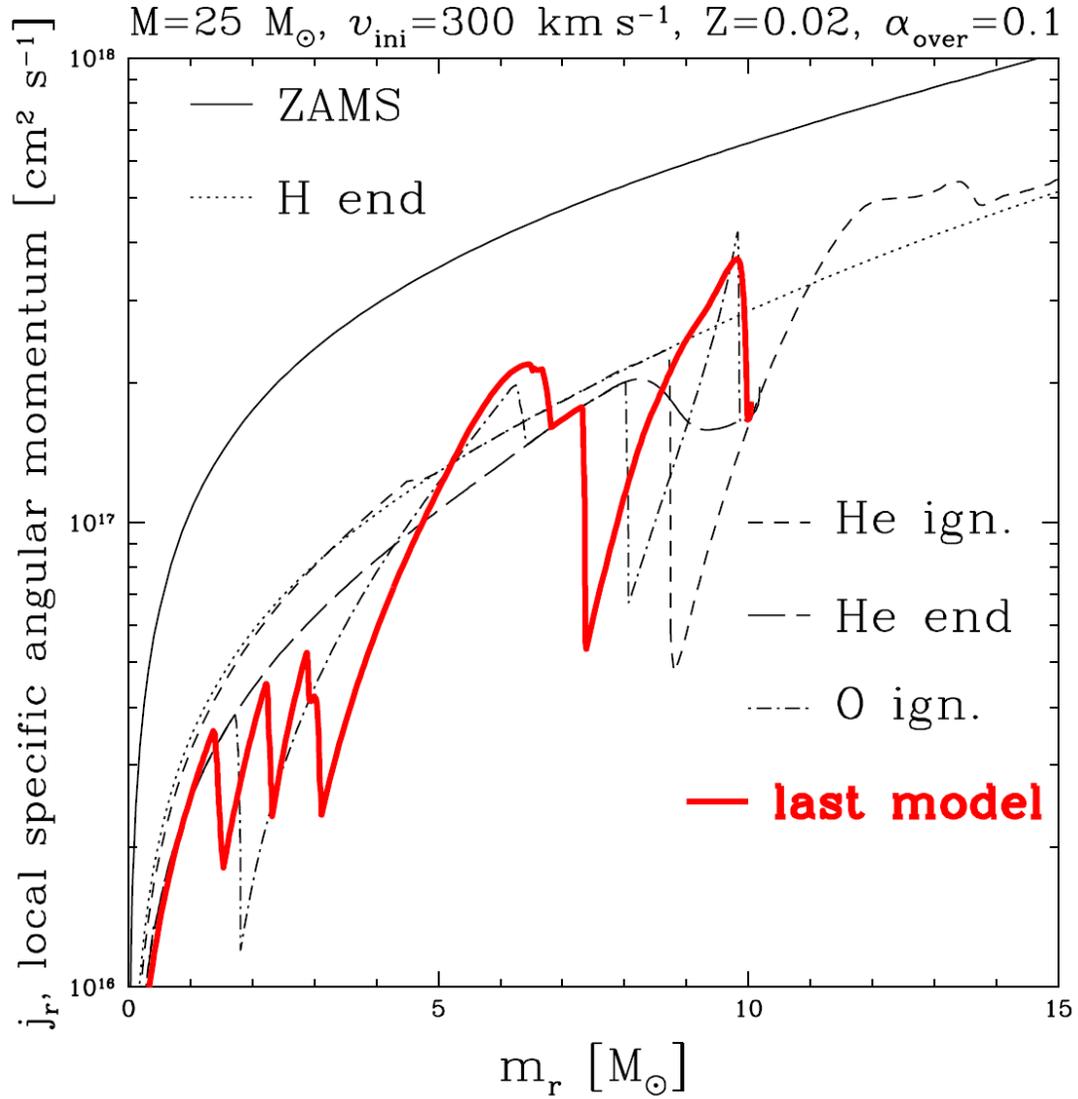

# FIG. 6

The internal distribution of the specific angular momentum at various evolutionary stages as a function of the Lagrangian mass coordinates inside a 25 Mo model with an initial rotation velocity of 300 km s$^{-1}$ . This is the same model as in Fig. 5, only the inner core of 15 Mo remaining at the end of the core Si-burning is represented. (From Hirschi et al. 2004).

# 4 EVOLUTIONARY EFFECTS OF ROTATION ON THE PROGENITORS AND THE TYPE OF EXPLOSIONS

The structure of the envelope determines the position of the core collapse supernova progenitor in the HR diagram and has a strong impact on the photometric and spectroscopic evolution of the supernova. In this section, we discuss how axial stellar rotation impacts the stellar envelopes and thus the nature of the progenitor in the HR diagram and the type of the SN event.

To discuss this topic we can distinguish four cases:
 i. The mass range of stars between 9 and 20 $M_{Sun}$: these stars end their life as red supergiants at solar metallicity (see for instance the 20 $M_{Sun}$ in Fig. 7). These stars will produce in general type IIP supernovae (see Filipenko 1997 for a discussion of the different types of core collapse supernovae).
 ii. The mass range between about 20 and 25 $M_{Sun}$: these stars cross the HR gap, being for a while a red supergiant, then they evolve back to the blue ending their lifetime as yellow, blue supergiants or even Wolf-Rayet stars. These stars are expected to produce type IIL, IIb supernovae in general and sometimes even type Ib (see the 25 $M_{Sun}$ in Fig. 7).
 iii. The stars in the mass range between 25 $M_{Sun}$ and 150 $M_{Sun}$: these stars end their life as WR stars. They may produce black hole with no supernova event, all the matter being swallowed by the stellar remnant, or a supernova event of type Ibc (see the tracks from 32 $M_{Sun}$ to 120 $M_{Sun}$ in Fig. 7).
 iv. Stars with initial masses above 150 $M_{Sun}$: these stars may encounter the pair instability during the advanced stages of their evolution. These instabilities can trigger pulsations and in some circumstances the complete destruction of the star giving birth to a Pair Creation Supernova (PCSN, Heger and Woosley, 2002).

The mass limits between these four categories are indicative and depend on many physical ingredients of the models like metallicity, rotation or stellar winds. Close binary interactions has also a strong impact on these issues.

4.1 Rotation and stars ending their lifetimes as red supergiants

Type IIP supernovae have since a long time been proposed to have red supergiants (RSG) as progenitors. To check that hypothesis, images of regions of the sky where type IIP supernovae appeared were examined in the archives of various instruments. In 13 cases, a star could indeed be detected (see Table 1 in Smartt, 2015) and in 13 cases no stars were detected, therefore only an upper limit for the luminosity of the progenitor could be deduced (see Table 2 in Smartt, 2015). These observations confirm the fact that type IIP supernovae have indeed red supergiants as progenitors. Progenitor masses between about 8 and 15-18 $M_{Sun}$ were obtained (Smartt, 2009). Interestingly, the low mass limit corresponds to the highest initial mass of stars that produce a white dwarf at the end of their evolution (see e.g. Fig. 7 in Doherty et al. 2016).

Since red supergiants are observed with masses well above 18 $M_{Sun}$, why don't we see progenitors of type II SNe with higher mass? At least two possibilities can be invoked here: first, it has been suggested that the more massive stars that end their evolution as red supergiants would produce a black hole without any visible explosion (Smartt 2009). A second possibility would be that red supergiants of higher masses are most of the time not the end point of the evolution. Due to high mass loss rates, stars with masses above about 18 $M_{Sun}$ would evolve back to the yellow-blue region of the HR diagram before exploding (see e.g. Georgy 2012, Meynet et al. 2015, Gordon et al. 2016). In that case, they would explode as Type IIL, Type IIn, Type IIb or even Type Ib supernovae. In Figs 7 and 8, evolutionary tracks for non-rotating and rotating stellar models at solar metallicity are shown (Georgy et al. 2012). We see that the upper luminosity for stars ending their lifetimes as red supergiants is between 20 and 25 $M_{Sun}$ for non-rotating stellar models. For the rotating models, it is between 15 and 20 $M_{Sun}$ quite well in agreement with the limit given by Smartt (2009) based on the mass estimates of type IIP progenitors. (Note that in Fig. 8 the 15 $M_{Sun}$ model is not shown. This star ends its lifetime as a red supergiant). Since stars have not all the same initial rotation, one expects to see diverse types of supernovae for a given initial mass.

4.2 The stars ending their lifetime as yellow-blue supergiants

A few cases of core collapse progenitors have been observed to be yellow or blue supergiants (see Fig. 9, see also Table 3 and Fig. 3 in Smartt 2015). The type of supernova associated with this kind of event is in general SN type IIL, IIb. These kinds of progenitors appear in the mass range between 15-20 $M_{Sun}$, thus near the upper limit given by Smartt (2009) for type IIP supernova progenitors. Interestingly the supernova light curve of the event that gave birth to Cas A could be analysed thanks to light echoes and it was found to be a type IIb supernova of initial mass around 15 $M_{Sun}$ (Krause et al. 2008).

Such progenitors can be produced by strong mass losses occurring during the red supergiant stage. These mass losses can be triggered either by some internal processes driven for instance by pulsations, or by interactions of radiation with dust. Also such yellow/blue supergiants may be produced by mass transfer in a close binary system. In that case the mass transfer does not necessarily occur during the red supergiant phase. It can occur in earlier phases.

In the same range of luminosities where these blue-yellow supergiant progenitors of supernovae are observed, red supergiants progenitors of type IIP supernovae are observed (see Fig. 9), indicating that in the same mass range, there is a diversity of evolution. What can be the physical reason for this diversity? In the frame of single star scenarios, axial rotation could be one factor. Indeed rotation enhances the luminosity of the star and thus favours the apparition of zones in the outer layers of the star where the luminosity becomes supra-Eddington, possibly triggering more efficient mass loss. The presence of a companion can also be such a factor.

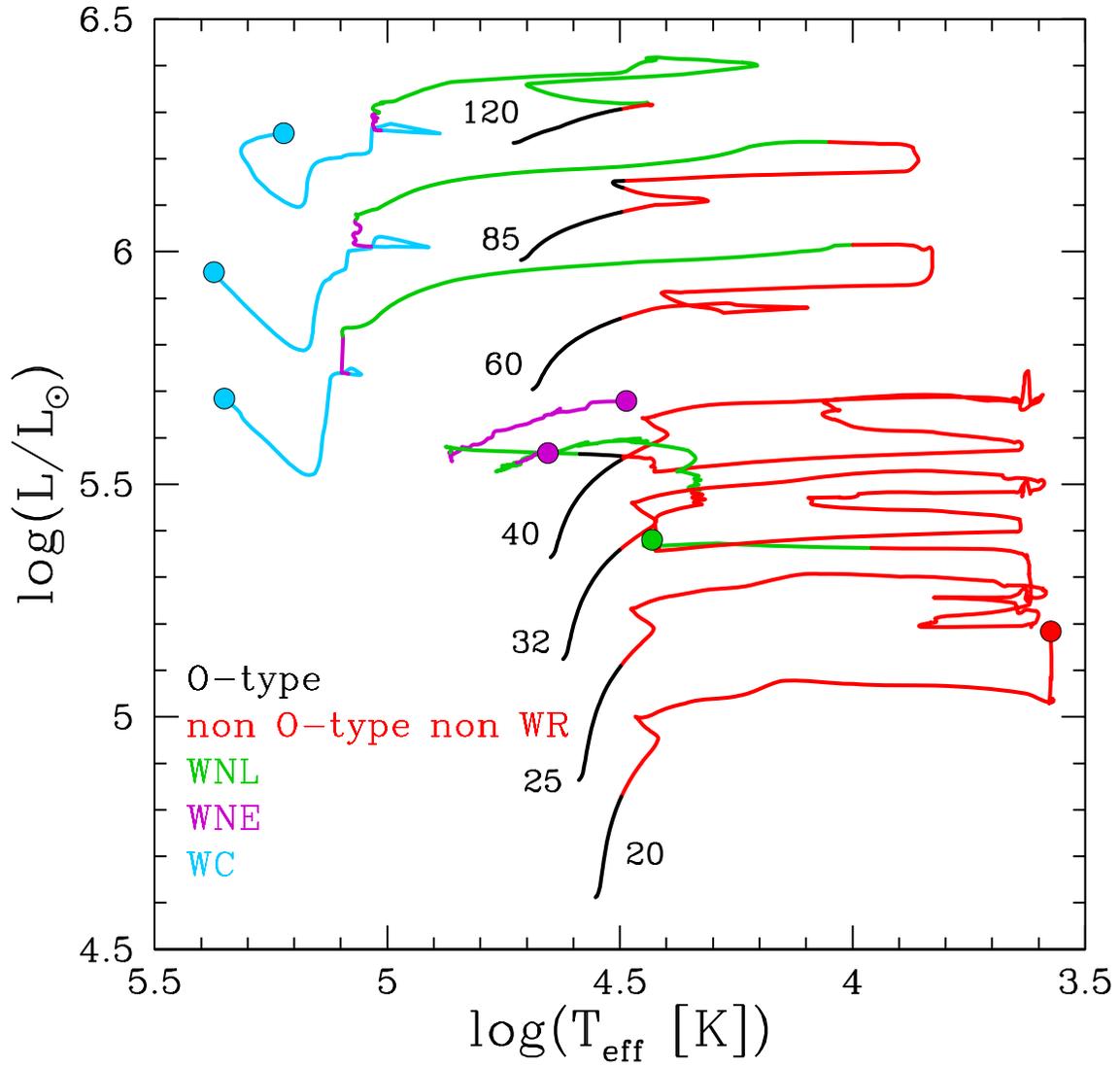

**FIG. 7**

Evolutionary tracks for non-rotating solar metallicity stellar models with initial masses above 20 $M_\odot$. The end point of the evolution is indicated by a circle. (From Georgy et al. 2012).

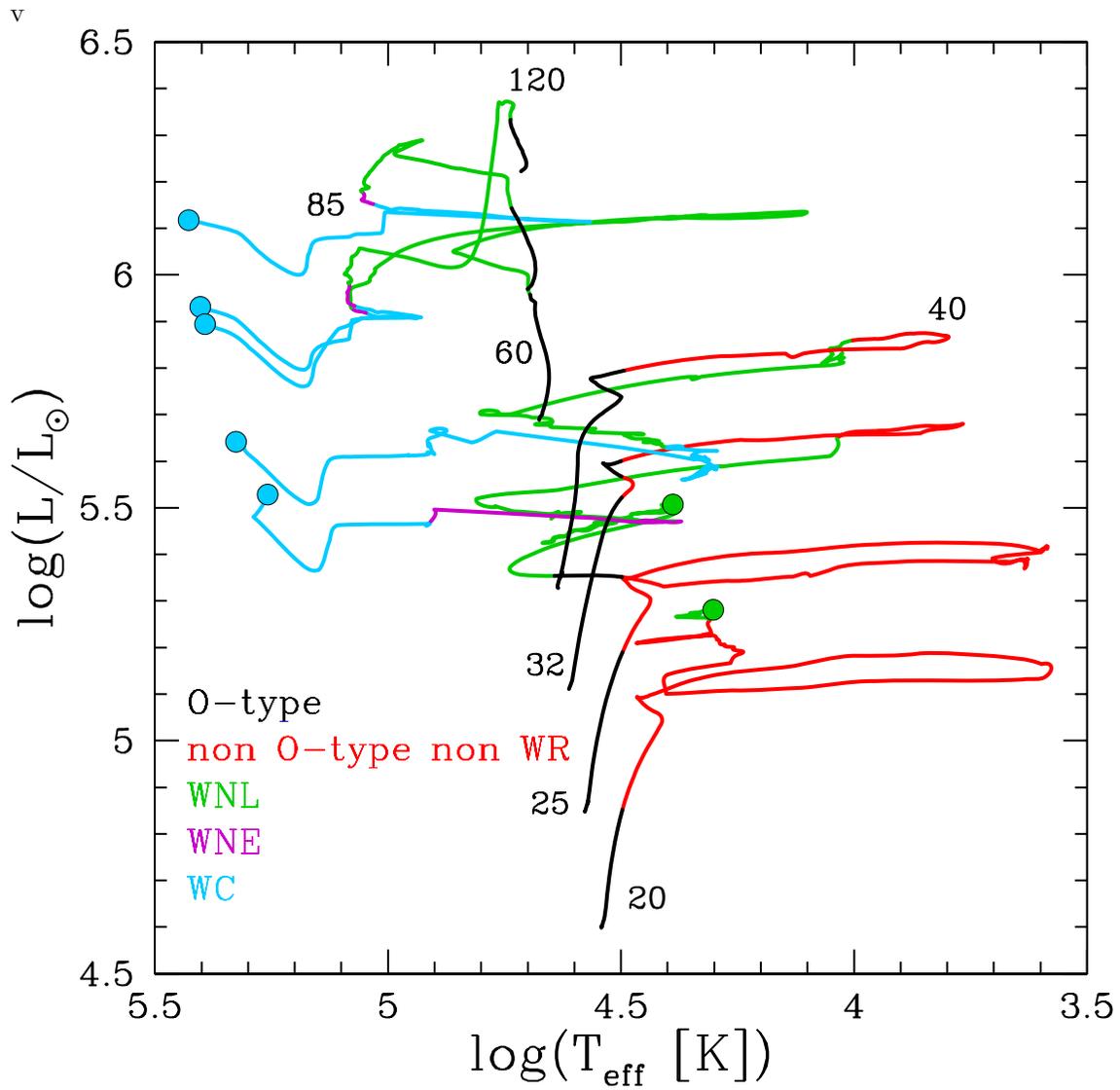

**FIG. 8**

Evolutionary tracks for rotating solar metallicity stellar models with initial masses above 20 $M_\odot$. The end point of the evolution is indicated by a circle. The rotating models correspond to an initial surface rotation at the equator equal to 40% the critical velocity (From Georgy et al. 2012).

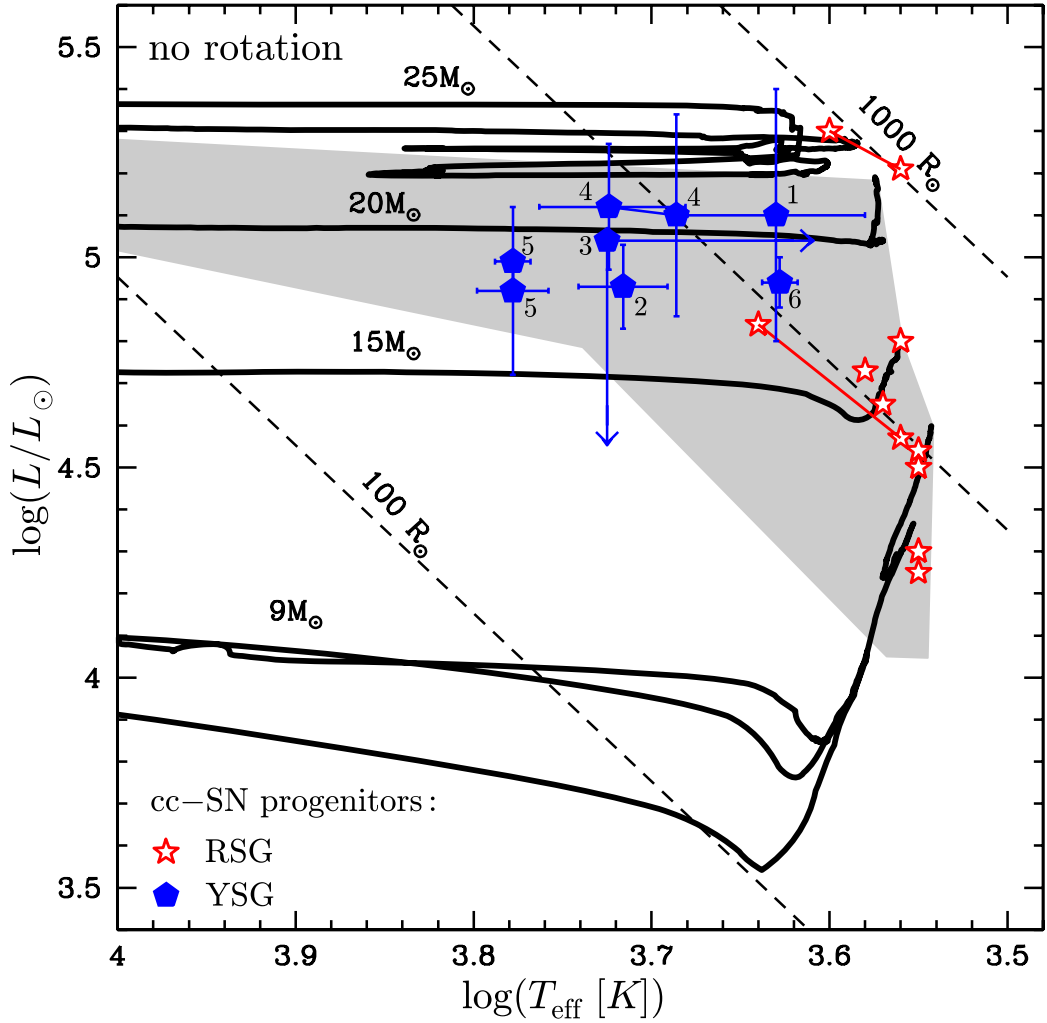

**FIG. 9**

Evolutionary tracks in the HR diagram for non-rotating models with superposed the positions of progenitors of core-collapse supernovae. Empty stars indicate red supergiants, filled pentagons those with a yellow supergiant as progenitors (see Table 6 in Meynet et al. 2015 for the references). A continuous segment links the positions of the same SN progenitors obtained by various authors. Progenitors of supernovae predicted by the nonrotating models computed with the various RSG mass-loss rates are found in the shaded area. Progenitors obtained with standard RSG mass-loss rates occupy the upper part as well as the right part of the shaded region, while the progenitors obtained from enhanced RSG mass-loss rate models are in the lower-left region of the shaded area. Lines of constant radius are indicated, the line intermediate between the 100 and 1000 R$_\odot$ corresponds to a radius of 500 R$_\odot$. Figure taken from Meynet et al. (2015).

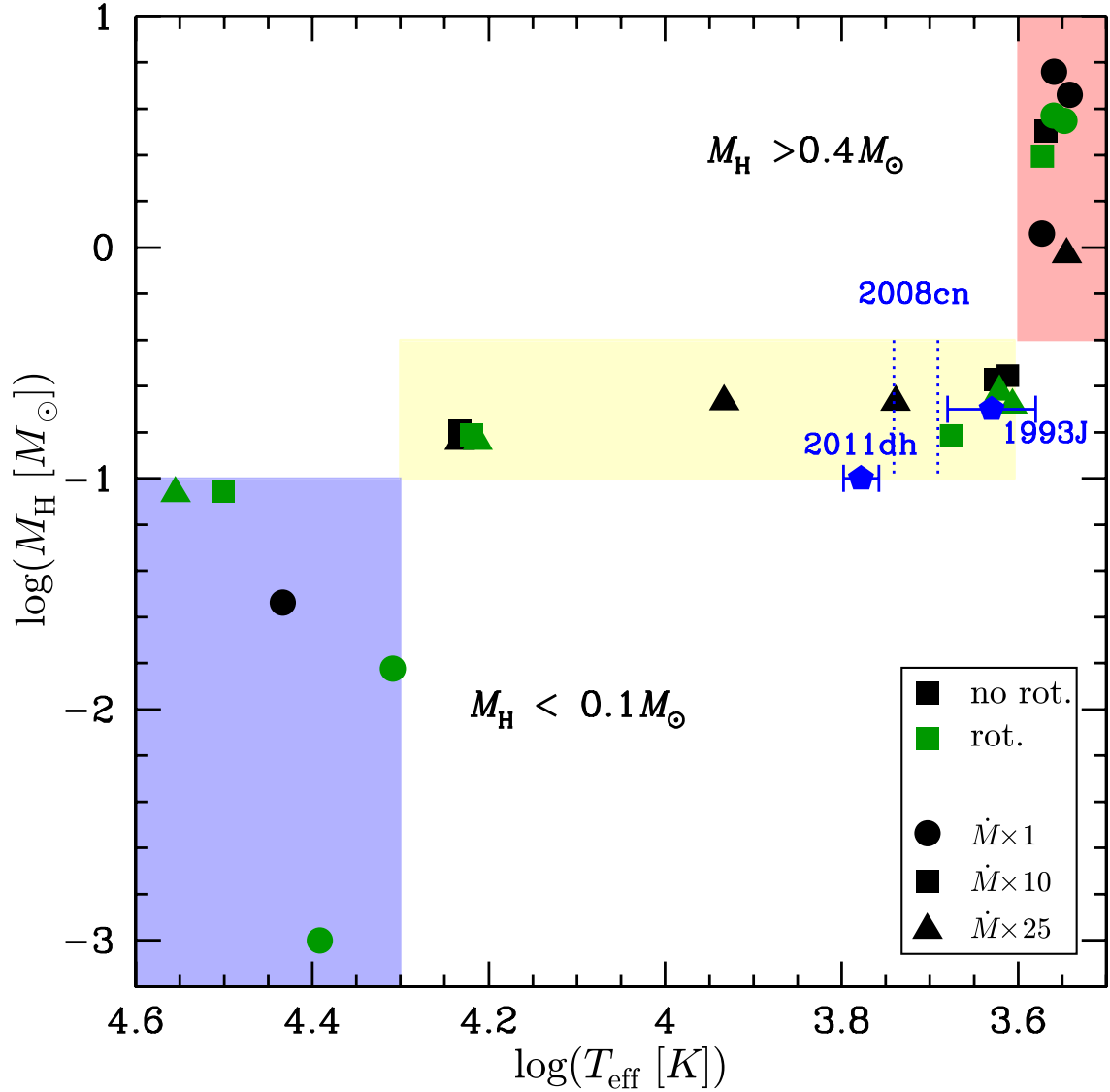

**FIG. 10**

Mass of hydrogen in solar masses at the pre-supernova stage for the various models computed with different mass loss rates during the red supergiant stage (initial masses between 9 and 25 Mo, see Meynet et al. 2015). Positions in this diagram of some supernovae are indicated by pentagons with error bars. For SN2008cn, the mass of hydrogen is not known, the range of effective temperatures is framed by the two dotted vertical segments. The three shaded regions from left to right correspond to stars ending their lifetime as blue or Wolf-Rayet stars, yellow, and red supergiants. Figure taken from Meynet et al. (2015).

Are there any possible expected signatures making it possible to distinguish between the single and the binary scenarios for making yellow-blue progenitors? In the case of the single star scenario, an evolution to the blue occurs when the core mass fraction becomes larger than a given limit around 70% of the total mass. This means that the yellow or blue supergiant will still have an H-rich envelope that contains about 30% of the total mass when it begins to evolve back to the blue. In case of a binary scenario, the mechanisms of the mass loss are governed by the evolution of the orbits of the two stars and the evolution of their radii. The mass transfer stops when the radius of the primary becomes less than the Roche limit. Interestingly, depending on the physical mechanism for the mass loss rates, the structure of the star might be quite different at the pre-supernova stage. Numerical experiments show that mass transfer can produce stars with a much smaller H-rich envelope than single star evolution. This difference might be a way to distinguish between these two scenarios in case some observations can be used to infer the mass of the H-rich outer layers in the progenitors.

In Fig. 10 the mass of H that remains in the envelope of various presupernova models is shown as a function of the effective temperature. These results were obtained from single star evolution models using different mass loss prescriptions during the red supergiant stage (Meynet et al., 2015). In order to obtain progenitor's effective temperatures between (in logarithm) 3.6 and 4.1, the mass of hydrogen in the envelope should be between 0.1 and 0.4 $M_{Sun}$. Estimates for H-mass can be obtained from the analysis of SN light curves. For the few cases where such estimates can be done, we see that they are well inside the limits given by the models. Figure 11 presents the spectral energy distribution and some information about the structure of some selected SN progenitor models. Passing from the red supergiant to the W-R star WO stage, the H-rich envelope decreases in mass and the progenitor decreases in radius. There is evidently a continuity in the amount of hydrogen left in stellar envelopes, which also implies a continuity in the associated types of supernovae from IIp, IIL to IIb.

4.3 The stars ending their lifetime as LBV stars

Luminous Blue Variables (LBV) have generally been considered as a transitory evolutionary stage between an O-type star phase and a Wolf-Rayet phase, or between a WNL-phase and a WNE or WC phase (see Crowther, 2007). Theoretical stellar models show that stars with initial masses larger than about 30-40 $M_{Sun}$ and with solar metallicity never become red supergiants (see Figs 7 and 8). These stars, generally at the beginning of their core He-burning phase, encounter some strong instabilities after the main sequence and lose in very short timescales an enormous amount of mass. One speaks in that case of eruptions rather than steady mass-loss episodes. Typically eta Carinae, which is probably the most studied LBV, lost in the middle of the 19$^{th}$ century in a period of 10-20 years about 10-20 times the mass of the Sun. See Humphreys and Martin (2012) for a recent discussion. The physical mechanism responsible for these eruptions is still not yet fully understood.

In general, these very heavy mass losses will rapidly cause the star to evolve away from the region where such processes are observed to occur, making it difficult to associate this evolutionary stage with a presupernova stage. Therefore, it was a surprise when some supernova events were associated with the explosion of an LBV, for instance recently SN

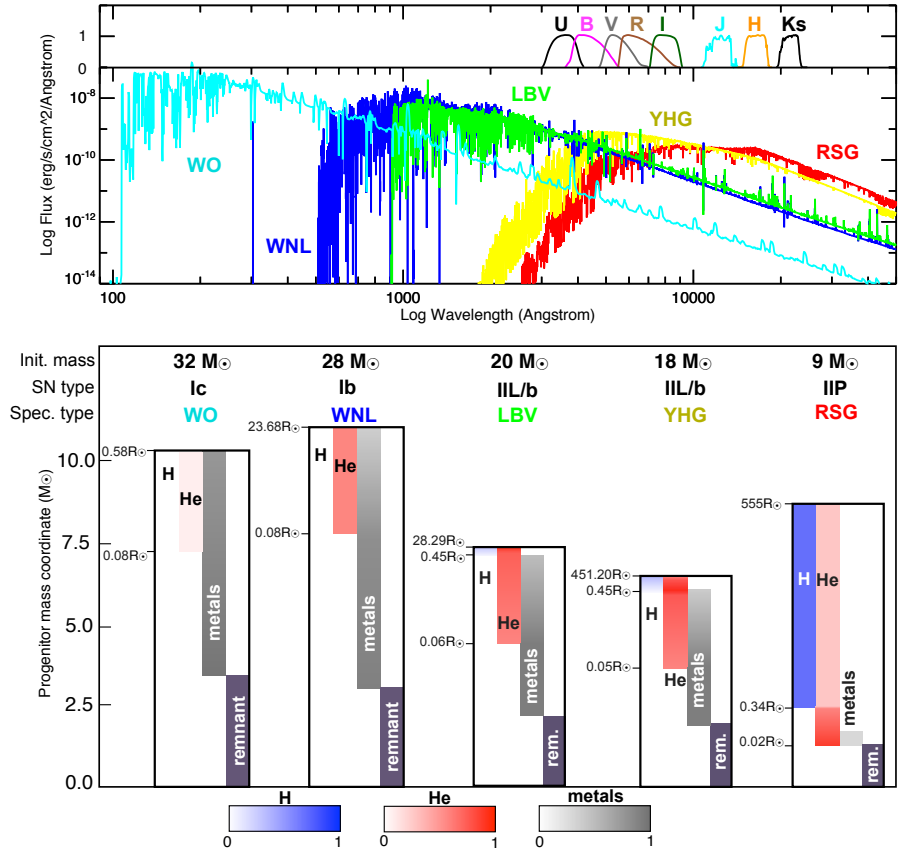

# FIG. 11

*Top:* Spectral energy distribution of selected models, showing a WO star with $T_{eff}$ 154000 K (32 M⊙ rotating model; cyan), a WNL star of spectral type WN10–11 with $T_{eff}$ 26800 K (28 M⊙ rotating model; blue), and an LBV with $T_{eff}$ 20000 K (20 M⊙ rotating model; green ). MARCS model spectra of a RSG with $T_{eff}$ = 3600 K and luminosity scaled to $L = 1.2 \times 10^5$ L⊙ (red), and of a YHG with $T_{eff}$ = 5250 K and luminosity scaled to $L = 1.5 \times 10^5$ L⊙ (yellow). All fluxes have been arbitrarily scaled to a distance of 1 kpc. The upper inset shows the normalized bandpasses of the *UBVRIJHKS* filters. We see that in a given filter, the most luminous star is not necessarily the one having the largest bolometric luminosity. *Bottom:* Schematic illustration of the interior structure of the SN progenitors for which spectra are shown in the upper panel. We show the Lagrangian mass coordinate of the progenitor in the y axis and the extension of the layers for different chemical elements (H, He, metals) and the baryonic remnant mass (computed as in Hirschi et al. 2005). The radius of selected shells is indicated on the left side of each sub panel, while the spectral type, SN type, and initial mass of the progenitors are indicated immediately above the interior structure. The chemical abundances (by mass) are color-coded in blue (H), red (He), and grey (metals). Figure taken from Groh et al. (2013c).

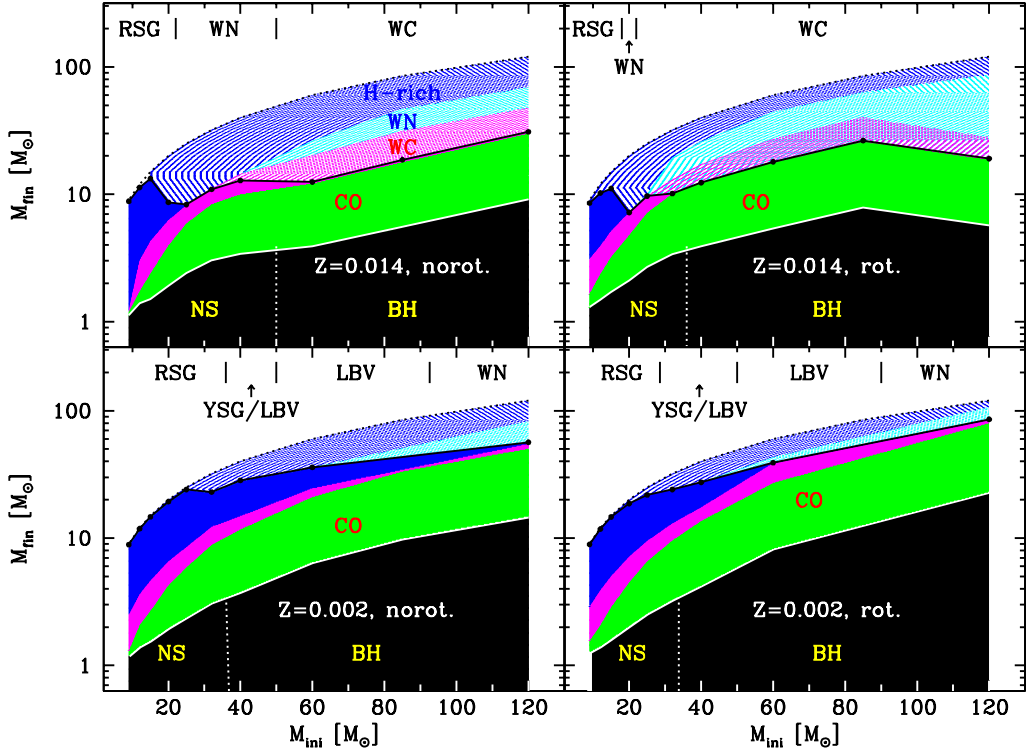

# FIG. 12

Masses of the stellar remnants and under the form of supernova and wind ejecta are shown as a function of the initial mass of stars for rotating and non-rotating models at solar metallicity and at the SMC metallicity. The masses that remains locked into the black hole (BH) or neutron star (NS) correspond to the solid black region. The masses ejected at the time of the supernova, rich in carbon and oxygen (solid green regions), in H-burning processed material (solid magenta regions), and ejected with the same abundances as the initial one (solid blue regions) are indicated. The hatched zones are the masses ejected by stellar winds. The blue hatched zone are H-rich stellar winds ejected before the entry of the star into the Wolf-Rayet regime. The cyan and magenta hatched regions correspond to the wind ejected masses when the star is respectively a WN and a WC star. The upper labels indicate the type of the core collapse progenitors. Models of Ekström et al (2012) and of Georgy et al. (2013) have been used.

2014C (Tinyanont et al., 2016). In those SN-LBV events, the SN light curve indicates that the explosive event occurred in a dense circumstellar environment typical of the one expected around a star having undergone LBV kind eruptions just prior the explosion.

Present day single star models can produce such occurrences in different ways. At solar metallicity, typically a 20 $M_{Sun}$ star may end its evolution in a blue region of the HR diagram where stellar winds are expected to show a bistability regime, i.e. an abrupt increase when some limit in effective temperature is crossed. It has been shown that the expected spectrum of such a star would be similar to the one of an LBV (although of low luminosity, see Groh et al. 2013a). Also it is expected that some radio emission should be observed when such a supernova occurs. This radio emission is synchrotron radiation emitted when the supernova ejecta interact with the progenitor's wind material (Moriya et al. 2013).

Another channel for producing an SN from an LBV progenitor is to consider high initial mass single stars in a low metallicity environment. Indeed, at high metallicities, the mass loss rates are so strong that an LBV is formed quite early in the core He-burning phase and there is no chance for the star to remain in that stage until the end of its evolution. It would evolve into a Wolf-Rayet stage (see Figs 7 and 8 and the upper panels of Fig. 12). Only at lower metallicities, due to much weaker mass losses, is it possible for the star to end its evolution as an LBV (see the lower panels of Fig. 12). For this to occur, however, the rotation should initially not be too strong otherwise it would induce a homogeneous evolution and the star would skip the LBV phase. Thus LBV SN progenitors in the high mass star range require both mass losses and internal mixing that are not too strong!

4.4 The stars ending their lifetime as Wolf-Rayet stars

Wolf-Rayet stars (WR stars) are defined as stars showing strong, broad emission lines in their spectra (see Crowther 2007 for a review on the physical properties of Wolf-Rayet stars). These emission lines are produced in hot circumstellar material expanding outwards at high velocities (typically 2000 km s$^{-1}$). Different subtypes of WR stars are observed, those presenting strong emission lines of helium and nitrogen (WN stars) and those presenting strong emission lines of helium, carbon and oxygen (WC, WO stars). The Wolf-Rayet phenomena can be produced in the central stars of planetary nebula and in massive stars with strong mass loss. Here we focus on this last category.

The origin of the Wolf-Rayet stars is a topic discussed for many decades but still some important questions remain. What are the evolutionary channels producing WR stars? More precisely what is the role of the single and binary channels in producing these stars? How can these different channels explain the observed variations of populations of WR stars in environments at different metallicities? What is the evolutionary link between the different subtypes?

Whatever the evolutionary scenario, it should produce a naked stellar core, that is a star which has lost most of its original H-rich envelope, exposing at the surface deeper layers where nuclear reactions (either the CNO cycle in the case of WN stars or He-burning reactions in case of WC-WO stars) have been active in the past. To make nuclear burning products appear at the surface of the star, there are two main possibilities that are not mutually exclusive and can work together: either stellar winds remove the H-rich envelope, exposing deeper layers at the surface, or internal mixing allows products of nuclear burning to diffuse up through the envelope and to appear at the surface.

Present day stellar models accounting for the effects of rotation and taking into account the most recent mass loss rate recipes, can produce a significant fraction (at least up to half) of the observed WR population in the Galaxy. Such models also predict that the number ratios of WR to O-type stars increase when the metallicity increases, in agreement with the observed trend. This is a consequence of the fact that stellar winds are stronger when the outer layers of the stars are richer in heavy elements.

Wolf-Rayet stars can also be formed via strong mass losses triggered by a Roche lobe overflow (RLOF) process. The largest differences with the single star scenario can be seen for stellar models that without the help of a companion would not succeed losing their H-rich envelope. For instance, such a channel might allow the formation of WR stars from relatively low initial mass stars and/or from stars, even massive ones, in metal-poor environments. Due to this last consideration, it was often argued that all the WR stars in the Small Magellanic Cloud (SMC) might well have been formed from close binary evolution. So it was surprising to realize that, as Moffat (2008) wrote: "the binary frequency of WR stars does not appear to vary with ambient metallicity, implying that RLOF is not important in most WR stars and that the presence of a massive star in a binary does not enhance its probability of becoming a WR star".

It might be also that a star observed as being isolated today has been part of a close binary system in the past. The companion has disappeared either because it has merged with the star observed today or because the system has been disrupted, for instance when the progenitor of the star that has now disappeared has exploded in a supernova. An asymmetric explosion may induce a kick that disrupts the binary leaving the visible star alone. A system born from a merging should likely show a high rotation unless some very efficient braking mechanism reduces the orbital angular momentum. The second scenario would produce runaway stars. At the moment, the observed population of WR stars does not seem to show frequently either evidence for a fast rotation or evidence for a fast proper motion, indicating that these two scenarios are unlikely to be the dominant ones.

When a WR star explodes, one expects that a type Ib (WN) or a type Ic (WC) supernova event occurs. If single stars produce these two types of supernovae at the end of their evolution, the models succeed in providing reasonable fit to the observed frequency of such supernovae within a factor 2 for non-rotating models and up to 84% of the observed value for rotating models (see Fig. 13 and 14). A different conclusion was obtained by Eldridge et al. (2013). These authors found that single WR stars contribute only one-fifth of type Ibc SNe, with binary systems giving rise to the rest. There are two arguments that would support the view that type Ibc SNe arise from close binary systems:

i) Models of SN light curves indicate that typically a mass of only 1-4 $M_{Sun}$ is ejected during type Ibc events (see e.g. Drout et al. 2011). Present day computations of single stars indicate

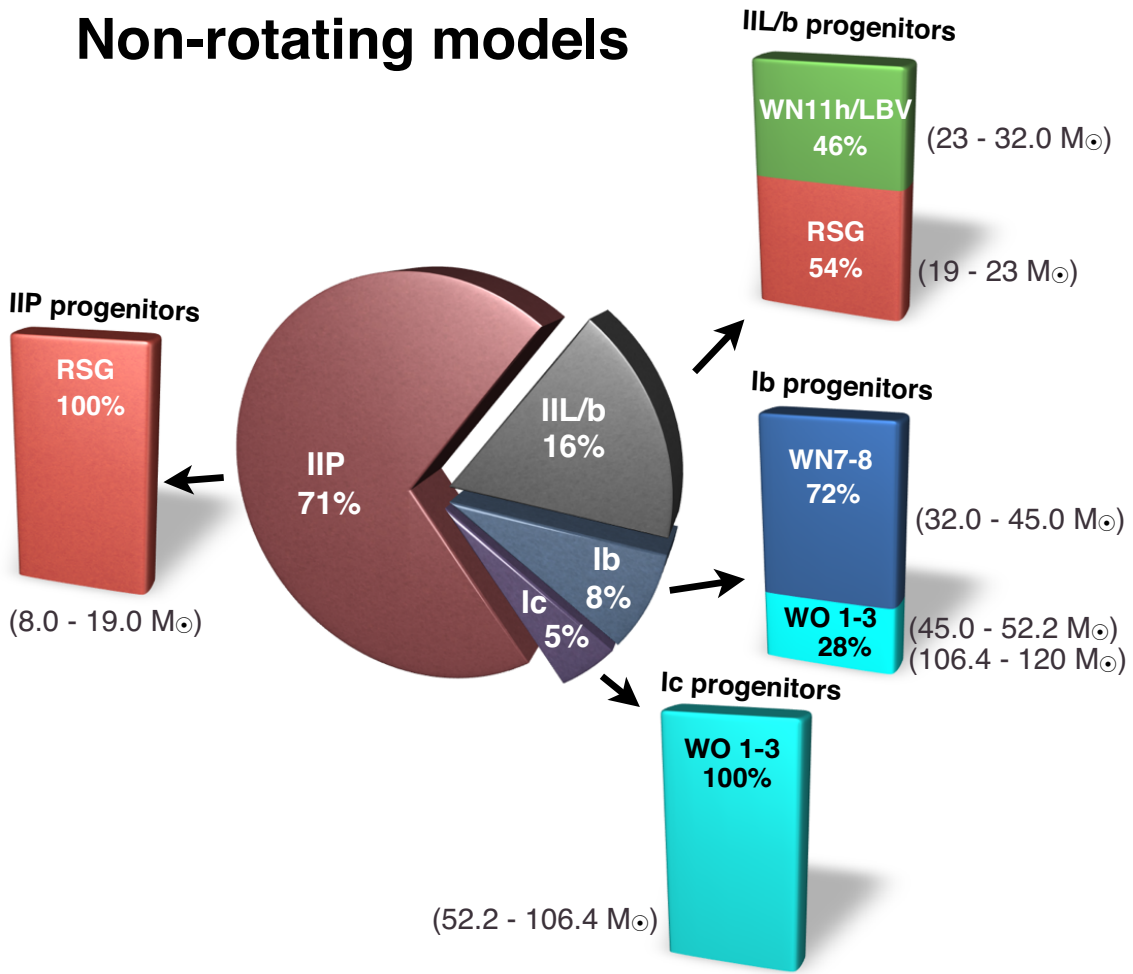

# FIG. 13

Diagram illustrating, for different core-collapse SN types, their relative rates and the types of progenitors and their respective frequencies as obtained from non-rotating solar metallicity models. Initial mass ranges (indicated in parenthesis) and SN types are based on the criteria outlined in Georgy et al. (2012), assuming that the minimum amount of He in the ejecta to produce a SN Ib is 0.6 Mo. According to Eldridge et al. (2013) the observed frequency are respectively about 56% (IIP), 15 (IIL/b), 9 (Ib), 17 (Ic). The remaining 3% are composed of 2% (IIn) and 1% (IIpec), Figure taken from Groh et al. (2013c).

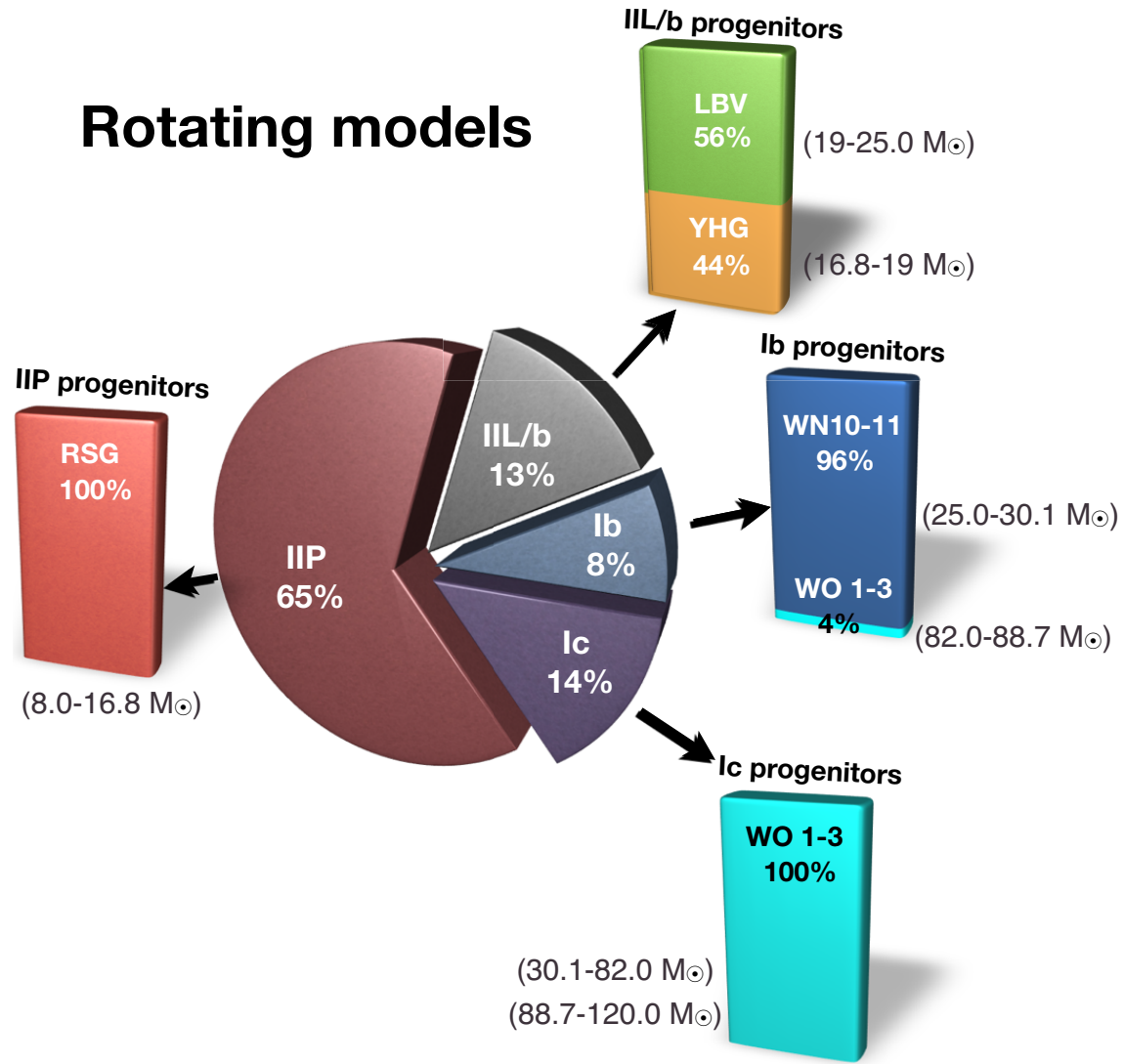

# FIG. 14

Same as Fig. 13 for solar metallicity rotating models. Figure taken from Groh et al. (2013c).

that the final masses are in the range of 10-15 $M_{Sun}$. If the explosion of such stars would produce 2 $M_{Sun}$ neutron stars, then, between 8 and 12 $M_{Sun}$ are expected to be ejected at the time of the supernova explosion, well above what is allowed by the study of type Ibc SN light curves. Unless much stronger mass losses are assumed, or there is stronger fall-back onto the stellar remnant, single star evolution cannot achieve the production of such low mass ejecta. Close binary evolution on the other hand could produce such small mass He-cores. The binary progenitor models from Eldridge et al. (2013) (which mainly form neutron-star remnants) have an average ejecta mass of $4.2 \pm 2.4$ $M_{Sun}$. To confirm such a scenario an important point to check is whether low luminous He-rich stars indeed exist. At the moment, such objects escaped detection. Maybe this is because they are produced only in the very late stage of their evolution and thus might be detectable only for a very short time. Another reason for their non-detection is that they might be outshone by their bright companion.

To conclude that argument, let us note that the small amount of ejected mass cannot be viewed as a too strong argument against single star models, because the final masses of stars are sensitive to the mass losses. We cannot be sure that the masses losses are correctly incorporated in the stellar models. If stronger mass losses than those presently used are appropriate then lower final masses could be achieved. Interestingly, at the end of their evolution, the WR star evolves near to their Eddington limit (see Fig. 5 in Maeder et al. 2012). This might trigger strong mass losses before the explosion and alleviate, in the frame of the single star scenario, the difficulty related to the mass of the supernova ejecta. Such strong mass losses, just prior the explosion, might also be an explanation for the non-detection of type Ibc progenitors (see below).

ii) Images of 14 regions where a type Ibc supernova occurred (and taken before the SN event) could be found in the archives of different instruments. In none of them (maybe with one exception, see below) could a progenitor be detected (Eldridge et al. 2013, see also Smartt 2015 for references to two more recent studies of type Ibc research of progenitors). This is puzzling because if WC stars are progenitors of such events, then at least in one case a progenitor should have been observed (Eldridge et al. 2013). This argument seems to indicate that the type Ibc events are not the result of the explosion of stars similar to the WR stars that are observed. Maybe this is due to the fact that the WR stars that are observed are likely not at their presupernova stage! Although the timescale until the end of the evolution can be short, it is not negligible (most, if not all, observed WR stars are likely in their core-He-burning phase). Predicting the colours of single WR presupernovae models from synthetic spectra, Groh et al. (2013c) showed that their magnitudes are actually below the observed upper limits given by Eldridge et al. (2013). Thus the non-detection does not appear at the moment to eliminate WR stars as the progenitors of this type of SNe.

There is one case of a possible detection of a progenitor for a type Ib SN, the case of PTF13bvn in NGC 5806, a galaxy at a distance of 22 Mpc (Cao et al. 2013). Cao et al. (2013) and Groh et al. (2013b) proposed that the broad band magnitudes of the source in the HST images were similar to massive WR stars. They proposed a progenitor of initial mass equal to about 30 $M_{Sun}$ evolved into a WN star, producing a final CO star mass of 10 $M_{Sun}$. Both Bersten et al. (2014) and Eldridge et al. (2015) presented binary models producing a stripped He-core with a final mass of 3−4 $M_{Sun}$. The ejecta mass would then be of order 2 $M_{Sun,}$ in agreement with the estimate of Fremling et al. (2014) from light-curve modelling.

It is likely that both channels, the single and the close binary one, contribute in making type Ibc supernovae. The respective importance of these two channels remains an open question that hopefully will be settled by improvements in both modelling and observations.

4.5 The stars ending their lifetimes as pair creation supernovae

The most massive stars encounter an instability in the advanced phases that might trigger pulsational instabilities and/or may lead to an explosion as a pair creation supernova (PCSN, Heger and Woosley, 2002, Woosley et al. 2007). This type of supernovae has been suggested to be associated with superluminous supernovae (Gal-Yam 2012).

The nucleosynthesis associated with these events shows remarkable features. According to Heger and Woosley (2002) who performed integration over a distribution of He-core masses, PCSNe produce a roughly solar distribution of even charged nuclei but are deficient in producing elements with odd nuclear charge. Also, essentially no elements heavier than zinc are produced.

Stars that enter into the advanced phases of the evolution with He-core masses between 64 and 133 $M_{Sun}$ produce PCSN. These stars reach such high temperatures in their central regions that the energy of the photons becomes sufficient to create pairs of electrons and positrons. This process of pair creation removes photons that play an important role in maintaining the mechanical equilibrium of the star. When occurring in a significant fraction of the star, the process leads to a collapse of the central region. This collapse accelerates the nuclear reactions and the rate of energy production is so large that the star is completely destroyed. Such events are very rare but since their whole material is dispersed they may contribute significantly to the chemical enrichment.

In order for such an evolution to occur, the star should not lose too much mass otherwise it would enter the advanced phases with a He-core mass below the minimum required for a PCSN to occur. This is the reason why such types of SNe are expected to appear more easily at low metallicities where the stellar winds are weaker. This can be seen in Figs. 12 and 15. According to the stellar models by Yusof et al. (2013), at solar metallicity none of the models is expected to explode as a PCSN. At the metallicity of the LMC, only stars with initial masses above 450 $M_{Sun}$ for the rotating models and above about 300 $M_{Sun}$ for the non-rotating case are expected to explode as a PCSN. At the SMC metallicity, the mass range for the PCSN progenitors is much more favourable. Yusof et al. (2013) obtain the result that all rotating stars in the mass range between about 100 $M_{Sun}$ and 290 $M_{Sun}$ would produce PCSNe.

Pop III stars are an interesting case. When no metals are present, one expects much more massive stars to be formed and thus more progenitors for PCSNe. However no traces of the peculiar nucleosynthesis of these stars have yet been found in long-lived, very metal-poor, low-mass stars formed from the ejecta of the first generations of massive stars. Reasons for this apparent absence could be multiple:
  i.   such massive stars might not be produced,
  ii.  their nucleosynthetic signatures should be found not in very metal poor stars but rather in stars with [Fe/H] around -2.5 because their iron yields are so large that stars produced from their ejecta should show a rather high metallicity (Karlsson et al. 2008),

iii. very massive Pop III stars are formed but they might lose, if they are rotating sufficiently fast, such a large amount of mass that they would avoid the Pair Creation instability.

This last case has been studied by Ekström et al (2008a) and is discussed in Sect. 6.

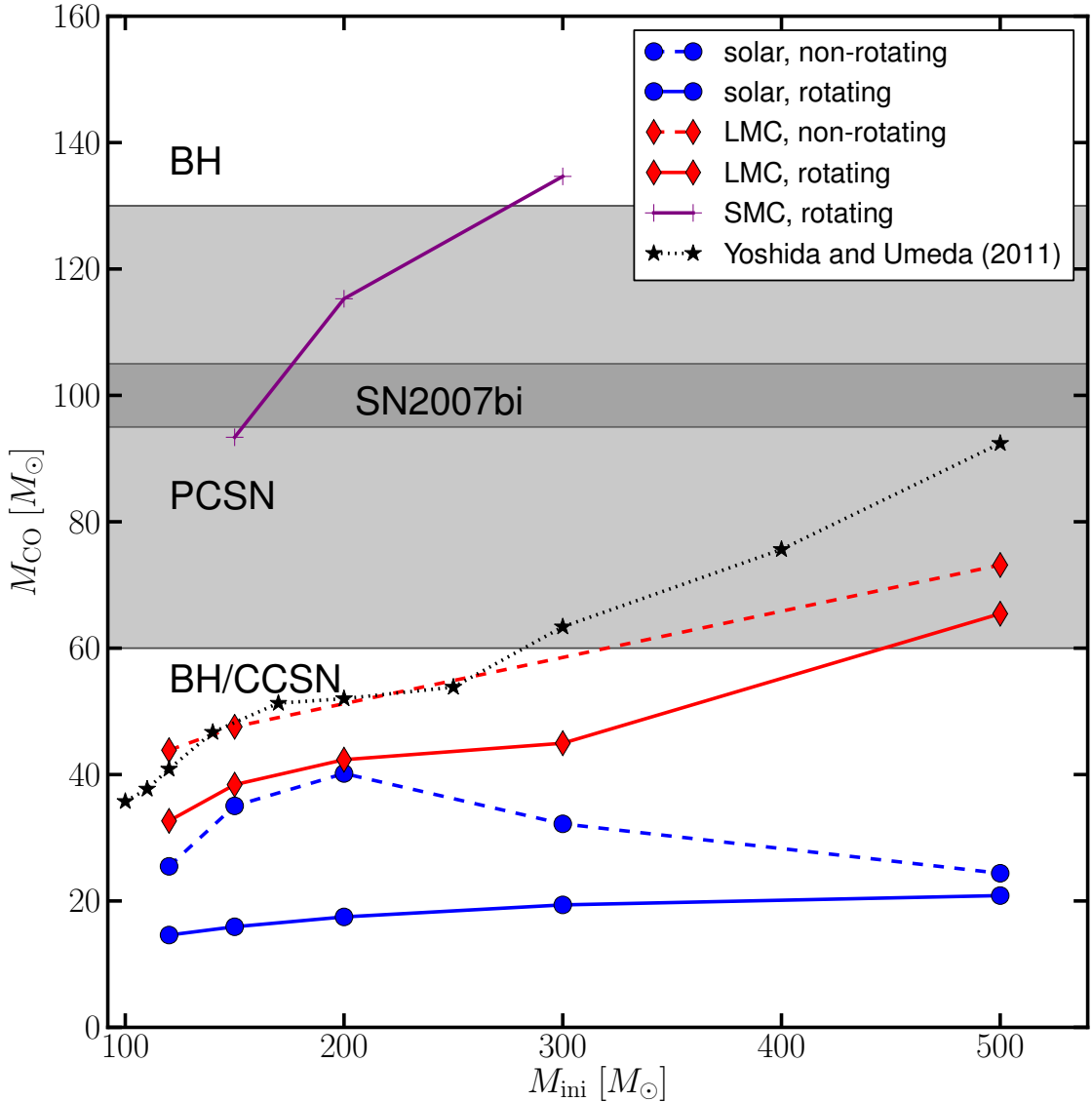

# FIG. 15

Masses of carbon-oxygen cores as a function of the initial masses for very massive stars at different metallicities. The light grey shaded area represents the range of $M_{CO}$, for which the estimated fate is a Pair Creation SN (PCSN). The thin dark grey shaded area corresponds to the estimated $M_{CO}$ of the progenitor of SN2007bi assuming it is a PCSN. The points linked by the dotted black line are from the models of Yoshida & Umeda (2011) at $Z = 0.004$, case A. Figure taken from Yusof et al. (2013).

## 5. THE CASE OF GAMMA RAY BURSTS AND MASSIVE BLACK HOLES

Gamma Ray Bursts (GRB) are the most energetic stellar explosions in the Universe. They are classified into two families, the long- and the short-duration GRB's. Here we shall focus on the long GRB that nowadays are considered to be produced by the explosion of massive fast rotating stars, while the short GRB are attributed to the process of merging of two neutron stars (see the review by Woosley and Bloom 2006).

The association, in a few cases, of a type Ibc supernova type to a long soft GRB provided some support to the collapsar model for GRBs proposed by Woosley (1993). In collapsar Woosley's model for GRBs it is assumed that a fast rotating black hole is formed. The fast rotation is important because it prevents the fall-back material being swallowed directly by the black hole. Instead the falling material forms an accretion disk. At least part of the gravitational energy released in the disk can be used to produce jets. Instabilities in these jets are invoked to produce the observed gamma rays. For those emissions to be visible, it is also needed that the progenitor of the black hole has lost most, if not all, its H-rich envelope. This means that, in the frame of this model, the progenitor is a Wolf-Rayet star.

One challenge of this model is to reconcile two conditions: fast rotation of the progenitor or at least of its central regions and the fact that this progenitor had to lose large amounts of material (and hence of angular momentum) to remove its H-rich envelope. There are at least two ways to circumvent this difficulty. One is to invoke not too strong coupling between the core and the envelope during the progenitor's lifetime. In that case the surface rotation may be very small (due to the high mass loss rate), and the rotation of the core remains quite high (since the coupling would be weak enough for preventing the core to be slowed down by the braking of the surface by the strong winds). This kind of evolution is predicted, for instance by shellular rotating models, where the coupling or the angular momentum transports are mediated through meridional currents and shear turbulence. In that case, indeed, even if the star loses large amount of material at the surface, initially fast rotating models on the ZAMS may still keep sufficient angular momentum in their core at the presupernova stage to fulfil the conditions for a collapsar (Hirschi et al. 2005b, Georgy et al. 2012).

When a strong coupling operates inside the star, then the strong mass losses needed to remove the H-rich envelope would break down the core in an efficient way and thus another kind of evolution is needed. One possibility is that very fast initial rotation induces such a strong internal mixing that the star would follow a homogeneous evolution (Yoon & Langer 2005). In that case, the removal of the H-rich envelope will be obtained through nuclear processing instead of mass loss (at least part of it, since mass loss will anyway also be present). This allows producing at the same time a fast rotating core and presupernovae models with no H-rich envelope. One key point in this scenario, however, is the mass loss rate that has to be used. Do these stars with such a peculiar evolution have mass loss rates similar to stars evolving in more standard ways (Maeder et al. 2012)? Homogeneous evolution implies also that the star will enter at an early stage of its evolution (actually during the Main Sequence phase) into the WR stage. It will thus undergo strong winds for a significant part of its lifetime. In order to preserve enough angular momentum in the core, a key point is to use WR stellar winds that are weaker than those used at solar metallicity, for instance. For a long time it has been argued that the WR mass loss rates do not depend on the initial metallicity. Actually, according to Vink and de Koter (2005, see also the review on stellar winds by Smith 2014) iron plays a key role in triggering mass loss in WR stars. So at low metallicities, it is expected that the WR stellar winds are weaker. In that case then, even if the star becomes a WR star due to internal mixing at an early stage of its evolution, homogeneous evolution can keep sufficient angular momentum in the core for the production of a collapsar.

The two categories of models - those with a weak internal coupling and those with a strong internal coupling - have their pros and cons. In the first case one may predict too many progenitors of GRBs, since the core can easily retain its angular momentum. This kind of model on the other hand might allow to explain the few GRBs that are observed at high metallicities, since these models somewhat decouple the evolution of the angular momentum in the core from that in the envelope. Strong internal coupling model on the other hand would produce favourable conditions for obtaining GRB only at low metallicities where the stellar winds are weaker and thus are less efficient in braking the star. Only the initially fast rotating massive stars at low metallicities are good candidates in that case.

The discussion above shows that in the framework of the collapsar model, the nature of the GRBs progenitors depends on the mass loss rates used for the stellar winds and still more importantly on the efficiency of the coupling between the core and the envelope.

Homogeneous evolution might also be an evolutionary channel leading to the formation of fast spinning close binary massive black holes (Mandel and de Mink 2016; Marchant et al. 2016). Some of these binary systems have characteristics at the end of their evolution for merging in a time inferior to the Hubble time and thus being sources of potentially observable gravitational waves (GW). According to Marchant et al. (2016, see their Fig. 9), sources of observable GW originating from homogeneous evolution would occur at low metallicities and would contain high mass and high spin black holes. The case of GW150914 (Abbott et al. 2016) might have been produced in that way.

## 6. THE FIRST STARS

Interestingly enough, there is an accumulation of theoretical and observational evidence which more and more supports the major role of rotation among the generations of first stars, with strictly $Z=0$ or extremely low metallicities. First, there are indications that suggest that stars rotate faster at lower Z. One was the finding that the relative frequency of Be stars strongly increases for lower Z, from the solar neighbourhood, to the Large and Small Magellanic Clouds (Maeder et al. 1999, Martayan et al. 2007). This does not mean in itself that this trend goes on towards $Z=0$. However, theoretical models of the formation of the first stars indicate that stars with very low Z should have very high rotational velocities, and thus experience mixing (Stacy et al. 2011).

CEMP stars are carbon enhanced metal poor stars, with [C/Fe] > +1 and [Fe/H] < -1. There are CEMP-r and CEMP-s stars, enriched in r- or s-elements. CEMP-no stars show no significant s-enhancement. CEMP-no stars have [Fe/H] between -2.5 and about -7, and with excesses of the [C/Fe] ratios up to 4 dex. Spinstars are low Z massive stars with fast rotation, strong mixing and high mass loss (Meynet et al. 2010). The interpretation in terms of spinstars of the huge abundance anomalies of the CEMP-no stars, which are the objects with the lowest known [Fe/H], supports the view that there are high rotational effects for the stars with the lowest metallicities. Models of massive stars with Z=0 and an initial rotation velocity equal to about half the critical value have been calculated by Ekström et al. (2008b). Other models have been calculated with $Z=10^{-8}$, $10^{-5}$ (Meynet et al. 2006, Hirschi 2007), $2\times10^{-3}$ (Georgy et al. 2013).

In stars of low Z, the effects of rotation, in particular internal mixing and mass loss, may be even stronger than at solar metallicity, thus favouring the formation of objects like spinstars. The stars are much more compact than at solar Z, because the opacity is weaker and thus radiation inflates the radii of the stars less. In the extreme case of stars with Z=0, the pp-chains are insufficient to sustain the luminosity and the stars further contract. For a given mass, their radii are about 3.5 times smaller than at solar metallicity. These smaller radii favour internal mixing since the mixing timescale goes like $R^2/D$, where D is the diffusion coefficient. Moreover the internal $\Omega$-gradients are higher and favour strong shear mixing, which has many consequences in the course of evolution.

Let us examine the evolution of spinstars. During the Main Sequence (MS) phase, the stellar cores are refuelled by internal mixing which brings fresh hydrogen. Thus, fast rotation makes the stellar cores bigger and the star more luminous. Such an effect is smaller in further stages, because the high gradient of mean molecular weight μ at the edge of the core tends to prevent strong mixing. Nevertheless, the core is initially larger and remains larger all the way up to the supernova stage. Another feature of the Z=0 and low Z models (present even at Z=0.002) is that the stars easily reach the break-up limit during their MS phase, even if their initial rotation velocity is moderate. Some mass is lost at this stage, but this remains of the order of a few percent of the total stellar mass.

The mixing in spinstars brings CNO products (mainly $^{14}$N and $^{13}$C) at the stellar surface. Also, C and O produced by core He-burning may be brought to the H-burning shell, thus producing primary $^{14}$N and $^{13}$C, further reactions involving the Ne-Na and Mg-Al cycle may also intervene. Primary $^{14}$N means nitrogen not made from the initial C and O (it would then be a secondary element), but from C and O produced in the star itself from the initial H and He elements. The surface enrichments in heavy elements like C increase the opacity in the outer layers and thus drive significant stellar winds (Meynet et al. 2006; Hirschi 2007). The mass loss by these winds can be very large particularly in the red supergiant stage, with the effect of reducing the final stellar mass and contributing to the chemical yields. Ejecta from stellar winds have much lower velocities than the ejecta from supernovae; thus they are more likely to contribute to the local chemical enrichments.

In the models at Z=0, the model with 150 M$_{Sun}$, a magnetic field and an initial velocity equal to half the critical value shows some interesting properties (Ekström et al. 2008a). The model rapidly reaches the critical velocity (the so called ΩΓ-limit if one takes into account the radiation pressure effect). The star model loses very little mass during the MS phase, but a lot in the red supergiant phase due to the chemical enrichment and keeping at the ΩΓ-limit. It enters the WR stage, characterized by further heavy mass loss. At the end of core He-burning, only 58 M$_{Sun}$ are remaining and this mass is lower than the minimum limit for a pair creation supernova (PCSN), which is 64 M$_{Sun}$ (Heger and Woosley 2002). Thus, rotation may allow the most massive stars to avoid the PCSN event. We note that the predicted nucleosynthetic signature of PCSN have not been observed up to now.

There is an impressive series of six recent kinds of observations which give support to the concept of spinstars in the early generations of massive stars (Chiappini 2013). These are:
 i. The evidences in the study of galactic chemical evolution for a production of primary nitrogen in stars with Z < 0.001.
 ii. In similar studies, the increase of the C/O ratios at low [Fe/H] also implies an injection of C by heavy mass loss.
 iii. The same trend is further supported by the low $^{12}C/^{13}C$ observed in metal poor giants of the galactic halo (Chiappini et al. 2008).
 iv. Another, however less direct, argument is based on the behaviour of the beryllium and boron [Be/H] ratios with respect to [O/H].
 v. The s-elements generally are accounted for by mass transfer in binary containing an AGB star. Such objects produce s-elements mainly of the second (Ba) and third (Pb) peaks, with very little elements of the first peak (Sr). The problem is that the observations of low Z stars show a large fraction of stars with a [Sr/Ba]>1, which cannot be produced by AGB stars. Most interestingly spinstars models (Frischknecht et al. 2016) show that the internal mixing in rotating massive stars produce significant amounts of s-elements, with much more from the first peak than the second or third one.

Another possible indication may come from the observations of a double sequence for many globular clusters (Piotto et al. 2005). This indicates that the relative helium to heavy elements enrichments ΔY/ΔZ is above 70, while supernovae only produce a ratio of about 4. A contribution from the winds of massive stars is one of the possibilities to explain such high ΔY/ΔZ ratios.

On top of these indications, the model of spinstars is well supported by the study of CEMP-no stars (Maeder et al. 2015). CEMP-no stars have not only low [Fe/H] ratios and large excesses in the [C/Fe] ratios, but also present a wide variety in the [C/Fe], [N/Fe], [O/Fe], [Na/Fe], [Mg/Fe], [Al/Fe], and [Sr/Fe] ratios. Back-and-forth motions with partial mixing between the He- and H-regions may account for this variety in the products of the CNO, Ne-Na and Mg-Al cycles. Some s-elements of the first peak may even be produced by these processes in a small fraction of the CEMP-no stars. Neither the yields of AGB stars (in binaries or not) nor the yields of classic supernovae can fully account for the observed abundance ratios in CEMP-no stars. Better agreement is obtained once the chemical contribution by stellar winds of fast-rotating massive stars is taken into account, where partial mixing takes place, leading to various amounts of CNO being ejected. These events occur before the corresponding supernova explosion, which will bring its contribution in heavy elements. However, the anomalous abundance ratios mentioned above appear to result from spinstar evolution before the supernova explosion.

Thus, on the whole there are really a number of facts in support of a strong rotational mixing in the early stellar generations at very low metallicities. The mass loss from these stars produces a significant nucleosynthetic contribution before the supernova explosions. These enrichments are particularly rich in products from the CNO burning, from the Ne-Na and Mg-Al cycles and from neutron captures producing s-elements of the first peak. All these effects imply that the role of rotation in the evolution of the first stars, and more generally among low metallicity stars, is very important and considerably influences the properties of the supernova progenitors, along the lines indicated in the above Sections 3 and 4.

REFERENCES

Abbott, B.P., and 974 colleagues 2016, Binary Black Hole Mergers in the First Advanced LIGO Observing Run. Physical Review X 6, 041015.

Bersten, M. C., Benvenuto, O. G., Folatelli, G., et al. 2014, iPTF13bvn: The First Evidence of a Binary Progenitor for a Type Ib Supernova, The Astronomical Journal 148, 68

Cao, Y., Kasliwal, M.M., Arcavi, I., et al. 2013, Discovery, Progenitor and Early Evolution of a Stripped Envelope Supernova iPTF13bvn, ApJL 775, L7

Chiappini, C., Ekström, S., Meynet, G. et al. 2008, A new imprint of fast rotators: low $^{12}C/^{13}C$ ratios in extremely metal-poor halo stars, A&A 479, L9

Chiappini, C. 2013, First stars and reionization: Spinstars, Astronomische Nachrichten, 334, 595

Crowther, P.A. 2007, Physical Properties of Wolf-Rayet Stars, Annual Review of Astronomy and Astrophysics 45, 177

De Mink, S. E., Cantiello, M., Langer, N. et al. 2009, Rotational mixing in massive binaries. Detached short-period systems, A&A 497, 243

Doherty, C. L., Gil-Pons, P., Siess, L., Lattanzio, J.C., Lau, H. 2015, Super- and massive AGB stars - IV. Final fates - initial-to-final mass relation, MNRAS 446, 2599

Drout, M.R., Soderberg, A.M., Gal-Yam, A., et al. 2011, The First Systematic Study of Type Ibc Supernova Multi-band Light Curves, ApJ 741, 97

Ekström, S., Meynet, G., Maeder, A. 2008a, Can Very Massive Stars Avoid Pair-Instability Supernovae? IAU Symposium, Volume 250, p. 209-216

Ekström, S., Meynet, G., Chiappini, C., Hirschi, R., &Maeder, A. 2008b, Effects of rotation on the evolution of primordial stars, A&A 489, 685

Ekström, S., Georgy, C., Eggenberger, P. 2012, Grids of stellar models with rotation. I. Models from 0.8 to 120 M at solar metallicity (Z = 0.014), A&A 537, A146

Eldridge, J. J., Fraser, M., Smartt, S. J., Maund, J. R., & Crockett, R. M. 2013, The death of massive stars - II. Observational constraints on the progenitors of Type Ibc supernovae, MNRAS 436, 774

Eldridge, J. J., Fraser, M., Maund, J. R., &Smartt, S. J. 2015, Possible binary progenitors for the Type Ib supernova iPTF13bvn, MNRAS 446, 2689

Filippenko, A.V. 1997, Optical Spectra of Supernovae, Annual Review of Astronomy and Astrophysics 35, 309

Frischknecht, U., Hirschi, R., Pignatari, M. et al. 2016, s-process production in rotating massive stars at solar and low metallicities, MNRAS 456, 1803

Fremling, C., Sollerman, J., Taddia, F., et al. 2014, The rise and fall of the Type Ib supernova iPTF13bvn. Not a massive Wolf-Rayet star, A&A 565, 114

Gal-Yam, A. 2012, Luminous Supernovae, Science 337, 927

Georgy, C. 2012, Yellow supergiants as supernova progenitors: an indication of strong mass loss for red supergiants? A&A 538, L8

Georgy, C., Ekström, S., Meynet, G. et al. 2012, Grids of stellar models with rotation. II. WR populations and supernovae/GRB progenitors at Z = 0.014, A&A 542, A29

Georgy, C., Ekström, S., Eggenberger, P., et al. 2013, Grids of stellar models with rotation. III. Models from 0.8 to 120 M at a metallicity Z = 0.002, A&A 558, A103

Gordon, M.S., Humphreys, R.M., Jones, T.J. 2016, Luminous and Variable Stars in M31 and M33. III. The Yellow and Red Supergiants and Post-Red Supergiant Evolution, arXiv:1603.08003

Groh, J. H., Meynet, G., Ekström, S. 2013a, Massive star evolution: luminous blue variables as unexpected supernova progenitors, A&A 550, L7

Groh, J. H., Georgy, C., Ekström, S. 2013b, Progenitors of supernova Ibc: a single Wolf-Rayet star as the possible progenitor of the SN Ib iPTF13bvn, A&A 558, L1

Groh, J. H., Meynet, G., Georgy, C., Ekström, S. 2013c, Fundamental properties of core-collapse supernova and GRB progenitors: predicting the look of massive stars before death, A&A 558, A131

Heger, A., Woosley, S.E. 2002, The Nucleosynthetic Signature of Population III, ApJ 567, 532

Heger, A., Langer, N., Woosley S.E. 2000, Presupernova Evolution of Rotating Massive Stars. I. Numerical Method and Evolution of the Internal Stellar Structure, ApJ 528, 368

Heger, A., Woosley, S.E., Spruit, H.C. 2005, Presupernova Evolution of Differentially Rotating Massive Stars Including Magnetic Fields, ApJ 626, 350

Hirschi, R. 2007, Very low-metallicity massive stars: Pre-SN evolution models and primary nitrogen production, A&A 461, 571

Hirschi, R., Meynet, G., Maeder, A. 2004, Stellar evolution with rotation. XII. Pre-supernova models, A&A 425, 649

Hirschi, R., Meynet, G., Maeder, A. 2005a, Yields of rotating stars at solar metallicity, A&A 433, 1013

Hirschi, R., Meynet, G., Maeder, A. 2005b, Stellar evolution with rotation. XIII. Predicted GRB rates at various Z, A&A 443, 581

Humphreys, R.M., Martin, J.C. 2012, Eta Carinae: From 1600 to the Present, Astrophysics and Space Science Library, Volume 384, 1

Karlsson, T., Johnson, J.L., Bromm, V. 2008, Uncovering the Chemical Signature of the First Stars in the Universe, ApJ 679, 6

Krause, O., Tanaka, M., Usuda, T., et al. 2008, Tycho Brahe's 1572 supernova as a standard typeIa as revealed by its light-echo spectrum, Nature 456, 617

Maeder, A. 2009, Physics, Formation and Evolution of Rotating Stars, Springer Verlag, 829 p.

Maeder, A., Meynet, G. 2012, Rotating massive stars: From first stars to gamma ray bursts, Rev. Modern Physics 84, 25

Maeder, A., Georgy, C., Meynet, G., Ekström, S. 2012, On the Eddington limit and Wolf-Rayet stars, A&A 539, A110

Maeder, A., Grebel, E. K., & Mermilliod, J.-C. 1999, Differences in the fractions of Be stars in galaxies, A&A 346, 459

Maeder, A., Meynet, G., Chiappini, C. 2015, The first stars: CEMP-no stars and signatures of spinstars, A&A 576, A56

Mandel, I., de Mink, S. E. 2016, Merging binary black holes formed through chemically homogeneous evolution in short-period stellar binaries. MNRAS 458, 2634.

Marchant, P., Langer, N., Podsiadlowski, P., Tauris, T. M., Moriya, T.~J. 2016, A new route towards merging massive black holes. A&A 588, A50.

Martayan, C., Floquet, M., Hubert, A. M., et al. 2007, Be stars and binaries in the field of the SMC open cluster NGC 330 with VLT-FLAMES, A&A 472, 577

Martins, F., Hervé, A., Bouret, J.-C. et al. 2015, The MiMeS survey of magnetism in massive stars: CNO surface abundances of Galactic O stars, A&A 575, A34

Meynet, G., Maeder, A. 1997, Stellar evolution with rotation. I. The computational method and the inhibiting effect of the μ-gradient, A&A 321, 465

Meynet, G., Maeder, A. 2002, The origin of primary nitrogen in galaxies, A&A 381, L25

Meynet, G., Ekström, S., & Maeder, A. 2006, The early star generations: the dominant effect of rotation on the CNO yields, A&A 447, 623

Meynet, G., Hirschi, R., Ekstrom, S., et al. 2010, Are C-rich ultra iron-poor stars also He-rich? A&A 521, A30

Meynet, G., Eggenberger, P., Maeder, A. 2011, Massive star models with magnetic braking, A&A 525, L11

Meynet, G., Chomienne, V., Ekström, S., et al. 2015, Impact of mass-loss on the evolution and pre-supernova properties of red supergiants, A&A 575, A60

Moriya, T., Groh, J.H., Meynet, G. 2013, Episodic modulations in supernova radio light curves from luminous blue variable supernova progenitor models, A&A 557, L2

Moffat, A.F.J. 2008, A Global Assessment of Wolf-Rayet Binaries in the Magellanic Clouds, Revista Mexicana de Astronomía y Astrofísica 33, 95

Piotto, G., Villanova, S., Bedin, L. R., et al. 2005, Metallicities on the Double Main Sequence of ω Centauri Imply Large Helium Enhancement, ApJ 621, 777

Przybilla, N., Firnstein, M., Nieva, M.F. et al. 2010, Mixing of CNO-cycled matter in massive stars, A&A 517, A38

Smartt, S.J. 2009, Progenitors of Core-Collapse Supernovae, Annual Review of Astronomy and Astrophysics 47, 63

Smartt, S.J. 2015, Observational Constraints on the Progenitors of Core-Collapse Supernovae: The Case for Missing High-Mass Stars, Publications of the Astro. Society of Australia 32, e016

Smith, N. 2014, Mass Loss: Its Effect on the Evolution and Fate of High-Mass Stars, Annual Review of Astronomy and Astrophysics 52, 487

Song, H. F., Maeder, A. Meynet, G. et al. 2013, Close-binary evolution. I. Tidally induced shear mixing in rotating binaries, A&A 556, A100

Song, H. F., Meynet, G., Maeder, A. et al. 2016, Massive star evolution in close binaries. Conditions for homogeneous chemical evolution, A&A 585, A120

Stacy, A., Bromm, V., & Loeb, A. 2011, Rotation speed of the first stars, MNRAS, 413, 543

Tinyanont, S., Kasliwal, M.M., Fox, O.D. et al. 2016, A Systematic Study of Mid-Infrared Emission from Core-Collapse Supernovae with SPIRITS, arXiv:1601.03440

Vink, J.S., de Koter, A. 2005, On the metallicity dependence of Wolf-Rayet winds, A&A 442, 587

Von Zeipel, H. 1924, Radiative equilibrium of a double-star system with nearly spherical components, MNRAS 84, 665

Woosley, S.E. 1993, Gamma-ray bursts from stellar mass accretion disks around black holes, ApJ, 405, 273

Woosley, S.E., Bloom, J.S. 2006, The Supernova Gamma-Ray Burst Connection, Annual Review of Astronomy and Astrophysics 44, 507

Woosley, S.E., Blinnikov, S., Heger, A. 2007, Pulsational pair instability as an explanation for the most luminous supernovae, Nature 450, 390

Yoon, S.-C., Langer, N. 2005, Evolution of rapidly rotating metal-poor massive stars towards gamma-ray bursts, A&A 443, 643

Yoshida T., Umeda H., 2011, A progenitor for the extremely luminous Type Ic supernova 2007bi, MNRAS 412, L78

Yusof, N., Hirschi, R., Meynet, G., et al. 2013, Evolution and fate of very massive stars, MNRAS 433, 1114

Zahn, J.-P. 1992, Circulation and turbulence in rotating stars, A&A 265, 115